\documentclass[10pt,journal,compsoc]{IEEEtran}
\ifCLASSINFOpdf
\else
\fi
\usepackage{amsmath}
\DeclareMathOperator{\argmin}{argmin}
\DeclareMathOperator{\argmax}{argmax}
\usepackage{threeparttable}
\usepackage{algorithmic}
\usepackage{algorithm}
\usepackage[style=base]{caption}
\usepackage[style=base]{subcaption}
\usepackage{booktabs}
\usepackage{multirow}
\usepackage{hyperref}
\usepackage{amsfonts}
\usepackage{graphicx}
\usepackage{titlesec}
\titlespacing*{\section}{0pt}{0.2\baselineskip}{0.2\baselineskip}
\titlespacing*{\subsection}{0pt}{0.2\baselineskip}{0.2\baselineskip}
\titlespacing*{\subsubsection}{0pt}{0.1\baselineskip}{0.1\baselineskip}
\renewcommand{\paragraph}[1]{\vspace{0pt}\noindent\textbf{#1}}
\usepackage[nodisplayskipstretch]{setspace}

\begin{document}
%
\title{Semantics-preserving Reinforcement Learning Attack Against Graph Neural Networks\\ for Malware Detection}
%
%
%
%

\author{Lan Zhang;
        Peng Liu,~\IEEEmembership{IEEE member};
        Yoon-Ho Choi;
        Ping Chen
\IEEEcompsocitemizethanks{\IEEEcompsocthanksitem Lan Zhang and Peng Liu are with the Department
of Information Science and Technology, Penn State University, State College,
PA, 16801.\protect\\
Lan Zhang: lfz5092@psu.edu\\
Peng Liu: pxl20@psu.edu\\
\IEEEcompsocthanksitem Yoon-Ho Choi is with Pusan National University. Email: yhchoi@pusan.ac.kr\\
\IEEEcompsocthanksitem Ping Chen  is with Fudan University. Email: pchen@fudan.edu.cn}
\thanks{Manuscript received Nov 13, 2020; accepted Feb 12, 2022.}}
%
%

\markboth{IEEE Transactions on Dependable and Secure Computing}%
{Shell \MakeLowercase{\textit{et al.}}: Bare Demo of IEEEtran.cls for Computer Society Journals}
%



\IEEEtitleabstractindextext{%
\begin{abstract}

As an increasing number of deep-learning-based malware scanners have been proposed, the existing evasion techniques, including  code obfuscation and polymorphic malware, are found to be less effective. In this work, we propose a reinforcement learning based semantics-preserving (i.e. functionality-preserving) attack against black-box GNNs (Graph Neural Networks) for malware detection. The key factor of adversarial malware generation via semantic {\tt Nops} insertion is to select the appropriate semantic {\tt Nops} and their corresponding basic blocks. The proposed attack uses reinforcement learning to automatically make these ``how to select" decisions. To evaluate the attack, we have trained two kinds of GNNs  with three types (e.g., Backdoor, Trojan, and Virus) of Windows malware samples and various benign Windows programs. The evaluation results have shown that the proposed attack can achieve a significantly higher evasion rate than four baseline attacks, namely the binary diversification attack,  the semantics-preserving random instruction insertion attack,  the semantics-preserving accumulative instruction insertion attack, and the semantics-preserving gradient-based  instruction insertion attack.

\end{abstract}


\begin{IEEEkeywords}
Adversarial samples generation, Malware detection, Graph Neural Networks, Reinforcement Learning.
\end{IEEEkeywords}}

\maketitle

\IEEEdisplaynontitleabstractindextext

%
\IEEEpeerreviewmaketitle

\section{Introduction}
\label{sec:intro}

The detect-evade ``game” between {\bf malware scanners} (e.g., antivirus software) and {\bf malware writers} is a long-lasting theme in cybersecurity. With the emergence of public online malware scanning platforms such as VirusTotal, it has been reported in recent years that some in-development malware were found on VirusTotal before their major outbreaks. For example, the very first known LeakerLocker sample could data back to November 2016 when it was submitted to VirusTotal \cite{Leaker2016}, but not until July 2017 did security analysts find it widespread. The study conducted in \cite{YHY19} shows that many malware writers have been anonymously submitting in-development malware samples to online malware scanning platforms. Since such platforms are hosting a comprehensive collection of state-of-the-art malware scanners, the scan reports would provide malware writers with a best possible assessment. In addition, through a continuous submit-and-{\bf revise} process, an in-development malware sample can evolve in such a way that when it was released into the wild and known to public, {\bf no signature} was available and all the malware scanners running on VirusTotal mistakenly reported benign.  Since it usually takes more than six months to generate well-crafted signatures \cite{Braue2015}, the above-mentioned evasion strategy renders a very serious security threat to the society. 

In order to help solve this serious problem, new kinds of malware scanners (e.g., \cite{DeLaRosa2018, yan2019classifying}) have been proposed in recent years to complement the existing scanners, and a key characteristic of these new malware scanners is that they all leverage machine learning, especially deep-learning (DL) models, to detect malware. The main merits of DL-based malware scanners are reflected in the following comparison: (a) While signature-based malware scanners often require substantial manual effort to extract signatures, DL-based malware detection models are automatically trained; (b) While signature-based malware scanners usually require a significant amount of domain knowledge, DL-based malware detection models could be well trained with minimum domain knowledge; (c) Due to the generalization ability of DL-based malware detection models, a malware sample revised (by its malware writer) to avoid having any known signatures could still be detected. 

Since the  DL-based malware scanners could make 
the above-mentioned evasion strategy ineffective, 
the malware writers must answer three key questions. {\bf Question 1:} Are the current malware revising techniques (e.g. code obfuscation) still effective? If not, why? {\bf Question 2:} If the current malware revising techniques are no longer very effective, how to revise a malware sample in a new way? {\bf Question 3:} How to ensure that the new way of malware revising will not introduce too much extra burden?

Although how to evade DL-based malware detection models has been attracting an increasing interest in the research community, the existing works are still very limited in answering the three questions. Almost all the existing works (e.g., \cite{anderson2018learning, Sharif, Park2019}) focus on investigating how CNN (convolutional neural network) or RNN (recurrent neural network) detection models could be evaded by adversarial examples. Although these works have made substantial progress on answering Question 2, most of them could not answer Question 1 or Question 3 due to the following reasons. First, to avoid changing the {\bf malware semantics} (e.g. functionality of the malicious code), most existing works restrict the
 adversarial examples by only modifying certain metadata, 
 such as PE header metadata and Section metadata\cite{anderson2018learning}, or injecting some bytes \cite{Kreuk}.
Second, 
although code obfuscation techniques such as polymorphic malware have been playing a very dominant role in real-world to evade malware scanners, very few existing works study the relationships between code obfuscation and adversarial examples. 
Since byte-level data (e.g., a malware sample is largely viewed as a stream of bytes) are used, the trained CNN and RNN models are 
   largely a black box and very hard to be explained. 
Accordingly, the semantics of the generated adversarial examples are very hard to be linked with code obfuscation (unless the models are assumed to be white-box). 
Hence, it is very difficult to answer Question 3 and reduce the extra burden by piggybacking the generation of adversarial examples on code obfuscation.   
 
In this work, we seek to answer these questions in a systematic manner.  
Our {\bf new insight} is as follows: instead of using CNN or RNN models, Graph Neural Network (GNN) malware detection models (e.g.,  
\cite{zhang2018end, yan2019classifying}), which are as
effective as CNN and RNN models, 
can be leveraged to discover and understand the {\bf inherent relationships} among {\bf code obfuscation}, the dominant technology in malware revising, and adversarial examples, the dominant concept in evading a deep neural network.  These inherent relationships then enable us to answer the three key questions. To the best of our knowledge, this is the {\bf first} work investigating how GNN based detection models could be automatically evaded by revising the 
basic blocks of a malware sample in a {\bf semantics-preserving} way. 

Our insight is gained based on two observations. First, 
since Control Flow Graphs (CFG) play an essential role in both 
code obfuscation and training of GNN based detection 
models \cite{yan2019classifying}, CFGs are a natural connection 
between code obfuscation and adversarial examples. 
Second, since CFG and basic block data are used to train 
a GNN based detection model, some adversarial examples 
are inherently related to basic-block-level code obfuscation. 
To systematically answer the aforementioned key questions, 
we face the following challenges:
1) Given a malware sample, the features in its graph structure involve discrete data, so we cannot directly leverage gradient-based attacks to 
generate infinitesimal small perturbation to its original CFG. 
2) The generated adversarial examples should not change the original 
malware semantics. In each adversarial example, which 
is also a malware sample, the attacker must
preserve exactly the same functionality as the original 
malware sample. Thus, the manipulation should not remove original features including edges and nodes.

To address these challenges, we proposed a novel method to automatically 
generate adversarial examples for GNN malware detection models 
while preserving the original malware functionality and semantics. 
We designed a reinforcement learning approach, namely the 
Semantics-preserving Reinforcement Learning (SRL) attack, to generate adversarial examples. 
The key factor of adversarial malware generation via semantic {\tt Nops} injection is to select the appropriate semantic {\tt Nops} and their corresponding basic blocks. However, since the decision making process involves discrete values, we cannot directly apply gradient based attack to make such decisions. 
The proposed SRL attack uses reinforcement learning 
to automatically make the above-mentioned decisions. 
These decisions result in 
 sequentially injecting semantic {\tt Nops} into the CFGs. 
Since semantic {\tt Nops} will never change malware functionality, 
the SRL attack achieves semantics-preserving. 

To evaluate the proposed SRL attack, we let it 
be compared with four baseline attacks: the binary diversification attack\cite{Sharif}, the Semantics-preserving Random Insertion (SRI) attack inspired by classical code obfuscation methods \cite{you2010malware}, the Semantics-preserving Accumulated Insertion (SAI) attack inspired by hill-climbing methods \cite{Sharif}, and the Semantics-preserving Gradient based Insertion (SGI) attack inspired by FGSM attacks\cite{goodfellow2014explaining}. 
For this purpose, 
we extracted CFGs from 8,000 benign and malicious real-world Windows 
programs and constructed abstract directed graphs to represent 
the CFGs. We then trained two kinds of GNN models, the basic GCN model \cite{kipf2016semi} and the DGCNN model\cite{zhang2018end}, respectively, 
to classify the graphs.

We found that the four baseline attacks are limited in evading 
GNN malware detection models. 
The experimental results have shown that the binary diversification attack achieved less than 10\% on the basic GCN models and the DGCNN models.
The SRI attack achieved an evasion rate of 45\% on the basic GCN models and 72\% on the DGCNN models. 
The SGI attack fooled the basic GCN model with an evasion rate near 41\% and deceived the DGCNN models with evasion rate 83\%. The SAI attack achieved 
over 90\% evasion rate on both models.  
In contrast, the proposed SRL attack achieved  
  100\% evasion rate on both the basic GCN models and the DGCNN models. 
We added the adversarial samples and retrained the detection models to 
defend against those attacks. The retrained models can achieve similar 
detection accuracy, but the evasion rates of the SRI attack, the SAI attack
, and the SGI attack significantly dropped 
to 2\%, 16\%, and 0.7\%, respectively. 
In contrast, the evasion rate of the proposed SRL attack 
dropped to 85\%, which is still a fairly high evasion rate. 

In summary, we have made the following contributions:
\paragraph{1)}
To the best of our knowledge, this is the first work on semantics-preserving black-box attacks against GNN malware detection models. The proposed SRL attack achieves semantics-preserving using reinforcement learning to select semantic {\tt Nops} and their corresponding basic blocks.
\paragraph{2)} Using both the basic GCN malware detection models and the DGCNN models,  
we evaluated and demonstrated the effectiveness of the proposed 
SRL attack via extensive experiments with 
different settings. 
We built a baseline with four attacks, and thoroughly compared
the proposed SRL attack with the baseline attacks. 
The results show that the proposed SRL attack is significantly more 
effective than the baseline attacks. 

The remaining of the paper is organized as follows. 
In Section \ref{sec:rw}, we review some background of malware detection and graph neural networks. The threat model and problem statement are presented in Section \ref{sec:problem}. 
The proposed SRL attack is described in Section \ref{sec:asg}. 
The proposed attack is compared with three baseline attacks in 
Section \ref{sec:exp}. In Section \ref{sec:related}, we discuss
the related works. Finally, we conclude the paper in Section \ref{sec:conc}.



\section{Background}\label{sec:rw}
In this section, we provide a background of the proposed attack.

\subsection{Malware Detection}

Malware detection models primarily make use of analysis techniques to understand the intention of malware. Features leveraged in malware detection can be grouped into three categories: static features, dynamic features, and hybrid features. Static features are extracted without running the executable files. Dynamic features are extracted by analyzing the behaviors of a program while it is being executed in a simulated and monitored environment. Hybrid features combine both static and dynamic features.

Various approaches are deployed to extract static features. Some of them make use of the binary file itself as indicators to detect the malware\cite{Cui2018}. 
The characteristics of the binary files, such as PE import features, metadata, and strings, are also ubiquitously applied in malware detection\cite{DeLaRosa2018}.
Others leverage reverse engineering, including instruction flow graph, control flow graphs, call graph, and opcode sequences\cite{DeLaRosa2018, xu2016hadm}, is used to understand the programs' architecture and extract related features.
Dynamic analysis executes the programs in a virtual environment to monitor their behaviors and observe their functionality.
Features obtained by dynamic analysis are API calls, system calls, registry changes, memory writes, network patterns, etc\cite{Tobiyama2016,Nix2017}.
Dynamic analysis can address some obfuscated malware and hence provide more accurate programs' behaviors. In line with the static analysis, attackers adopt approaches to prevent malware from dynamic analysis\cite{Ispoglou2016}. The malware starts an early check and immediately exits if it runs on virtual machines. Some malware even executes benign behaviors so humans draw incorrect conclusions about the intent of the malware.

Because the signature-based detection method is not resilient to slight variations, researchers have applied conventional machine learning algorithms (e.g., Random Forest) and deep learning method to detect malware. 
Convolutional neural networks and fully connected dense layers are leveraged to learn the high level features out of the selected features\cite{Cui2018}. For sequential data such as API calls and instruction sequence, recurrent neural networks (RNNs) are applied to classify the malware~\cite{Nix2017, Tobiyama2016}.
Recently some researchers use graph neural networks to classify malware programs represented as control flow graphs\cite{yan2019classifying}. The structure information contained by CFGs can be used to find unreachable code, find syntactic structure (like loops), and predict programs' defect\cite{Machiry2018,Phan}. 



\subsection{Graph Neural networks}
In this work, we study attacks targeting malware detection models built from CFG-represented data. 

To embed structural information inherent in graph like data, we use two graph neural network: the basic GCN model\cite{kipf2016semi} and the DGCNN model\cite{zhang2018end}. 
In alignment with the DGCNN model\cite{zhang2018end}, our graph convolution layer takes the following propagation rule:
\begin{equation}\label{equ:gcn}
    H_{l+1} = \sigma (\widetilde{D}^{-1}\widetilde{A}H_lW)
\end{equation}
Here, $\widetilde{A} = A+I$ is the adjacency matrix $A$ of the directed graph $G$ with added self-connections $I$. $\widetilde{D}_{ii} = \sum_{j}\widetilde{A}_ij$ is its diagonal degree matrix. $W$
is a layer-specific trainable weight matrix. $\sigma(\cdot)$ is a nonlinear activation function. $H_l$ is the matrix of activations in the $l$-th layer; $H_0=X$, where X denotes the node information matrix of graph $G$. 

The graph convolutional layer propagates node features to neighboring nodes as well as the node itself to extract local substructure information. We stack multiple graph convolution layers to get high-level substructure features. 
For the basic GCN model, we add a classification layer after node embedding to extract graph features. The classification layer simply flattens the high-level substructure features and adds a fully-connected layer followed by a nonlinear activation function.
For another model, we follow the same architectures as the DGCNN model \cite{zhang2018end}. First, we concatenate the output of multiple GCN layers. Then, we use the SortPooling layer to sort the features followed by 1-D convolutional layers and dense layers to learn the graph-level features.

\subsection{Reinforcement Learning}
As a decision-making algorithm, Reinforcement Learning (RL) applies to many real-life sequential decision-making tasks under uncertainty with the objective of optimizing reward. The RL tasks can be formed as a finite horizon markov decision process (MDP) that contains four components: 
1) discrete set of environment states $S$; 2) discrete set of environment actions $A$; 3) transition function $T$ that calculates the transition probabilities between states; 4) reward function $R$ that provides short-term rewards of the state. At each time step $t$, the agent takes an action $a_t$ from the action set $A$ for current state $s_t$ and the environment provides a reward $r(s_t,a_t)$. 
According to the transition probability and the reward, the state transitions to the next state $s_{(t+1)}$.
The goal of the RL agent is to learn a policy $\pi(s,a)\in\Pi $ to decide which action to perform in a given state, where $\Pi$ is a set of policies. Under a given policy $\pi$, the Q-value of an action $a$ with a state $s$ is
\begin{equation}
Q^\pi(s,a) = \mathbb{E}[r_1+\gamma r_2+...|s,a,\pi],
\end{equation}
where $\gamma$ is the discount factor that trades off the importance of immediate and future rewards.
Q-learning fits Bellman optimality equation for the Q-value function as follows:
\begin{equation}
 Y_t^Q= r(s_{t+1}, a_{t+1}) + \gamma \max_{a'\in A} Q^\pi(s_{t+1}, a';\theta_t),
\end{equation}
where $\theta_t$ refers to the parameters that define the Q-values at the $t$-th iteration. 
The deep Q-network (DQN) algorithm trains a neural network $Q(s,a;\theta)$ to select actions for each state. The DQN algorithm uses a target network, with parameters $\theta^-$ to update the target value:
\begin{equation}
 Y_t^Q= r(s_{t+1}, a_{t+1}) + \gamma \max_{a'\in A} Q^\pi(s_{t+1}, a';\theta_t^-),
\end{equation}
The  parameters $\theta^-$ are updated every C iterations with $\theta^-=\theta$. The experience replay in an online setting keeps past experience $(   s,a,r,s^{'}  ) $. Every $T$ time step, a mini-batch experience is selected within the replay memory and used to train the Q network. Here, the loss function to update parameters $\theta$ is the square loss:
\begin{equation}
L_{DQN}=(Q(s,a;\theta)-Y_t^Q )^2.
\end{equation}
\subsection{Code Obfuscation}
Code obfuscation tools serve two main purposes: (1) to protect intellectual properties; (b) to evade malware detection systems. There are a variety of code obfuscation schemes. A basic requirement is that the program semantics must be preserved after the code is transformed by such tools. Traditionally, attackers use obfuscation methods, including dead-code insertion, register reassignment, subroutine reordering, instruction substitution and so on, to morph their malware to evade malware detection\cite{you2010malware}. Here we list the definition of some obfuscation methods:
1) Semantic {\tt Nops} Insertion: inserting certain
ineffective instructions, e.g., $NOP$, to the original binary without changing its behavior;
2) Register Reassignment: switching registers while keeping the program code and its behavior same, such as registers $EAX$ in the binary are reassigned to $EBX$;
3) Instruction Substitution: replacing some instructions with other equivalent ones, for example, $xor$ can be replaced with $sub$;
4) Code Transposition: reordering the sequence of the instructions.


Code obfuscation is effective to evade the signature-based detection system because it could significantly change the syntactic of original malware. 
Although code obfuscation tools have enabled attackers to successfully evade various real-world malware detection systems, the unique capabilities of deep learning models, especially graph neural networks, indicate that the evasion ability of code obfuscation tools can no longer be taken for granted. 
The experiments in our paper demonstrate that randomly inserting dead instruction can not achieve a good result compared to other attacks with the help of prediction confidence.


\section{Problem Statement}\label{sec:problem}
To answer the overarching research question, "\textit{How effectively can we exploit the potential shortcomings of the deep learning malware detection models?}", we studied the resilience of GNN malware detection models over CFG-represented malware. After showing how to represent malware using CFG, we formulate the evasion problem based on the code obfuscation methods. 


\subsection{CFG-Represented Malware Model}\label{subsec:graph}
A CFG is a directed graph representation that illustrates all reachable paths of the program during execution. Nodes of the CFG represents the basic blocks of the program. Each basic block is a consecutive, single-entry code without any branching except at end of the sequence. Edges in the CFG represent possible control flow in the program. Control enters only at the beginning of the basic block and leaves only at the end of the basic block. Each basic block can have multiple incoming/outgoing edges. Each edge corresponds to a potential program execution. 
By using only opcodes such as $MOV$, $ADD$ or $JMP$ as features of the instruction after ignoring all operands, we abstract a basic block. For example, $sub\ \%eax,\%ebx$ and $sub\ [\%ecx],\%edx$ would be both represented by $sub$ in a basic block. This is because some instructions, such as $JMP$ and $INC$, share the same bytes, but they have different semantic meanings. Also, operands or instruction values result in large syntactic differences, e.g. different registers being used, but the semantic meaning of the instructions can be similar. 



From the executable files, we generate a CFG $G=<V, E>$, where $v_{i} \in V$ and $e_{ij} \in E$. Each node $v_i$ represents a basic block in a CFG, while a directed edge $e_{ij}$ points from the first basic block $v_i$ to the second basic block $v_j$. We also adopted the concepts of Bag-of-words (BoW) in NLP to vectorize the basic blocks. First, we map $n$ opcodes of the x86 instruction set to a list $S = \{s_1, s_2, ..., s_n\}$, where $s_i$ is a particular opcode. After counting the occurrence of opcodes in the basic block $v_i$, a basic block $v_i$ in a CFG $G=<V, E>$ is transformed as an array of integer counts with size same as $S$. Let $v_i = \{x_1...x_n\}$, a vector of counts over each opcode. $x_k$ is set as the occurrence if the $k$-th opcode exists in the basic block, otherwise, it is set as 0. \label{sec:encode}




In the context of the adversarial malware functionality, we can not directly change edges in the CFG to modify the adversarial malware. To preserve the functionality of the malware, we limit manipulation on the malware into semantic {\tt Nops} insertion. That is, when defining a list of semantic {\tt Nops} that will not affect the program functionally, we do not change the structure of the original CFG. 



\subsection{Problem Formula}\label{sec:nop}
In this paper, we consider the black-box setting where the attacker can only receive the final estimation(probability) results from the malware detection model $\mathbb{C}$. Even though a malicious CFG $G=<V, E>$ can be correctly labeled as a malware by the pretrained and fixed malware detection model, the attacker aims to safely manipulate the basic blocks and generate an adversarial graph $\widetilde{G}=<\widetilde{V}, E>$ to deceive the malware detection model $\mathbb{C}$.


Let $G=<V, E>$ be a given graph, where $V\in \mathbb{R}^{m\times n}$ and $E \in \mathbb{R}^{m\times m}$, $m$ is the number of nodes and $n$ is the number of opcodes in the instruction set. The semantic {\tt Nops} are encoded using the aforementioned method in Section \ref{sec:encode}. The manipulation on malware is small and imperceptible perturbations that should not change the program's original functionality. We define a list of dead instructions $\zeta \in \mathbb{R}^{I \times D}$, where $I$ is the number of dead instructions. By selecting and inserting the dead instructions into the original sample $G$, we generate an adversarial sample $\widetilde{G} = <V+\delta , E>$. The adversarial sample has the same graph structure as the original sample. The proposed semantics-preserving attacks aim to maximize the probability of the target label under the constraints that at most $\Delta$ instructions can be injected. Thus, an adversarial example is generated by solving the following constrained optimization problem:
\begin{equation}\label{equ:opt}
 \begin{aligned}
 \argmin_{\widetilde{G}} \quad & \mathbb{C}(\widetilde{G}, \widetilde{y}; \theta)\\
  \quad \textrm{s.t.} \quad & d(G, \widetilde{G}) \leq \Delta,\\
\end{aligned}
\end{equation}
where $\widetilde{y}$ is the target label, $\mathbb{C}(\cdot)$ is the malware detection model with parameters $\theta$, and $d(\cdot)$ is the distance function to calculate the number of injected instructions.


\section{Attacks on Graph-based Malware Detection Model}\label{sec:asg}
To solve the optimization problem in Eq.~\ref{equ:opt}, we designed four semantics-preserving attacks against GNN for malware detection. While one is a semantics-preserving reinforcement learning attack which is designed by sequentially injecting semantic {\tt Nops} into the CFG, the other three attacks are semantics-preserving attacks designed using the ideas such as random insertion, hill-climbing, gradient-based insertion methods respectively.


\subsection{SRL Attack}


The key factor of adversarial malware generation via semantic {\tt Nops} injection is to select the appropriate semantic {\tt Nops} and their corresponding basic blocks. However, since the decision making process involves discrete values, we can not directly apply gradient based attack to inject semantic {\tt Nops}. Reinforcement learning can be a valid approach to attack GNNs for malware detection when modifying the graph structure and node features~\cite{dai2018adversarial, sun2019node}. Thus, we design a semantics-preserving reinforcement learning attack which results in sequentially injecting semantic {\tt Nops} into the CFGs.

\begin{figure*}[t]
  \centering
  \includegraphics[width=\linewidth]{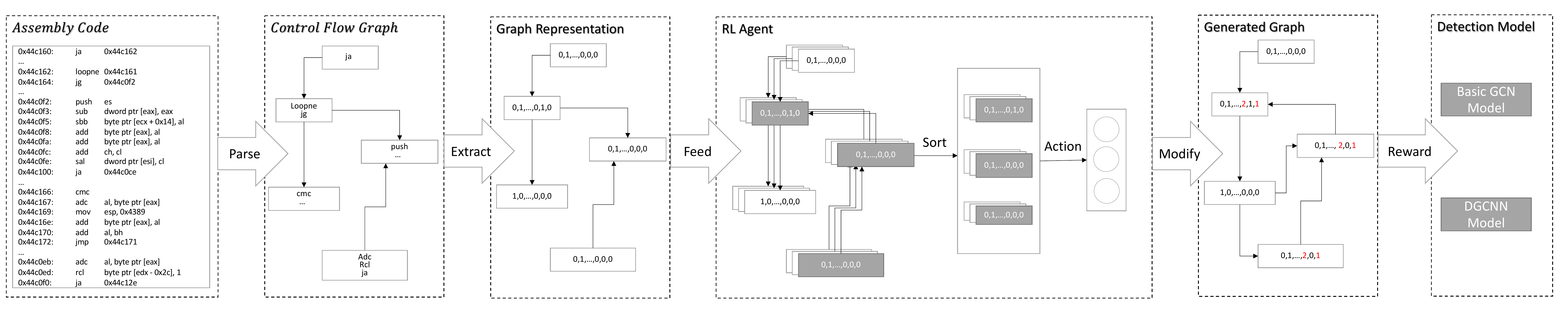}
  \caption{Architecture of the SRL Attack}
  \label{fig:arch2}
\end{figure*}

As shown in Figure \ref{fig:arch2}, the RL agent iteratively chooses basic blocks and semantic {\tt Nops} while modifying the generated malicious graph until it successfully misleads the malware detection model. During each iteration, the RL algorithm records the $topk$ basic blocks represented by embedding the opcodes into the CFGs as described in Section \ref{subsec:graph}, the semantic {\tt Nops} chosen by the RL agent and the rewards calculated by feeding the generated CFG to the detection model. 
The injected features do not affect the program functionally, and change the structure of the original CFG. 
Using a small batch of training records, the RL agent learns which semantic {\tt Nops} should be chosen and which basic blocks should be modified. As a result, the learned RL agent selects the modified basic blocks and the semantic {\tt Nops} for one CFG to maximize rewards and evade the detection model.
To solve the optimization problem in Eq.~\ref{equ:opt}, we design reinforcement learning environment and reward as follows:

\paragraph{State.}
The state $s_t$ at time $t$ represents a CFG $G_t=<V_t, E>$ with some of the manipulated basic blocks. 

\paragraph{Action.}
Each action includes two folds: 1) the importance of the basic blocks $v_t$; 2) a dead instruction $a_t$. The action space of picking up a dead instruction is $O(I)$, where $I$ is the numbers of semantic {\tt Nops}. Similar to \cite{Sharif}, below is the rule we followed when generating the semantic {\tt Nops}:
1) Some atomic instructions that do not change the memory or register value, e.g. $NOP$.
2) An invertible instruction, such as arithmetic operation and logical operation, followed by the inverse instruction, e.g. $PUSH$, $POP$, $ADD$ and $SUB$. 

\paragraph{Reward.}
The ultimate goal of the reinforcement learning model is to generate new samples that can misclassify the detection model. 
In practice, the decision process will take long to find the right action during training process. Thus, we calculate the rewards of each state as an intermediate feedback. 
If one CFG can successfully avoid detection, the reward $r_t$ is associated with the action sequence length.
We design the guiding reward $r_t$ 
to be one if it increases the probability $p_{s_t,a_t,v_t} = \mathbb{C}(s_t,a_t,v_t,\widetilde{y})$ of successfully evading the prediction model after being recognized as the target label $\widetilde{y}$, and to be zero otherwise. 
\begin{equation}\label{equ:reward}
    r_t(s_t, a_t, v_t) = \left\{\begin{matrix}
 1;& if \ \ \ p_{s_t,a_t,v_t} > p_{s_{t-1},a_{t-1},v_{t-1}} \\ 
 0;& otherwise.
\end{matrix}\right.
\end{equation}

\paragraph{Terminal.}
Once the injected instructions reach the budget $\Delta$, or current state can be misclassifed, the process stops. 

\begin{algorithm}[t]
\begin{algorithmic}[1]
\begin{scriptsize}
\renewcommand{\algorithmicrequire}{\textbf{Input:}}
 \renewcommand{\algorithmicensure}{\textbf{Output:}}
 \REQUIRE $\mathbb{C}(\cdot)$, $\widetilde{y}$, $NopsList$, $topk$, $niters$, $\Delta$, $T$, $C$
 \STATE Initialize $Q(s,a,v)$ with random parameters $\theta$\;
 \STATE Set target function $\hat{Q}$ with parameters $\theta^-=\theta$\;
 \STATE Initialize replay memory buffer $\mathcal{M}$\;
 \FOR{each $G=<V, E>$ }
 \STATE $t \leftarrow  0$ \;
 \STATE $s_t \leftarrow G$ \;
 \WHILE{$\argmax(\mathbb{C}(s_t)) \neq \widetilde{y}$ and $t < niters$}
 \STATE   With probability $\epsilon$ select a random action $a_t$ and $v_t$\;
 \STATE Otherwise select $a_t$ and $v_t$ by the Q network\;
 \STATE   Insert the action $a_t$ into $topk$ basic block $v_t$ to get new graph $G'$ as $s_{t+1}$\;
 \STATE   Compute $r_t$ by feeding $G'$ to $\mathbb{C}$\;
  \IF{$Diff(s_{t+1}, s_t) > \Delta$}
  \STATE   $t\leftarrow niter$
  \STATE $r_t\leftarrow 0$
  \ENDIF
  \STATE Calculate the absolute temporal difference(TD) error $|\delta_t|$\;
 \STATE   Store $\{s_t, a_t, v_t, r_t, s_{t+1}, |\delta_t|\}$ in memory $\mathcal{M}$\;
   \IF{Every $T$ queries}
   \STATE   Sample minibatch transitions $\{s_j, a_j, v_j, r_j, s_{j+1}\}$\;
   \STATE Calculate the target value and the square loss based on the target function $\hat{Q}$ and the reward $r_j$\;
   \STATE   Update parameters of the Q network \;
   \ENDIF
   \IF{Every $C$ queries}
   \STATE Reset $\hat{Q} = Q$
   \ENDIF
   \STATE $s_t\leftarrow s_{t+1}$
  \STATE   $t\leftarrow t+1$\;
 \ENDWHILE
 \ENDFOR
\end{scriptsize}
 \end{algorithmic}
 \caption{SRL attack against malware detection}
 \label{alg:srl}
\end{algorithm}



As other work on adversarial graph generation\cite{dai2018adversarial}, the policy network used to learn the MDP is Q-learning, specifically the deep Q-network (DQN) algorithm. 
In Algorithm~\ref{alg:srl}, we show the operational procedure of the DQN algorithm for generating the proposed SRL attack. 
The RL agent iteratively chooses basic blocks and dead instructions while modifying the malicious input until it successfully misleads the malware detection model.
First, a CFG is represented as a state $s_0$. Second, for each turn $t$, the RL agent chooses an action $a_t \in \mathcal{A}$ to decide which dead instruction should be inserted into the observable environmental state vector $s_t$. 
For each malicious CFG, the RL agent chooses $topk$ basic blocks based on the output of the SortPooling layer and one semantic {\tt Nop} in the list of semantic {\tt Nops}. 
To determine $topk$ basic blocks $v_t$ that will be manipulated at each iteration, the RL agent sorts the basic blocks in the graph according to their importance. 
The RL agent uses the SortPooling layer to calculate nodes' importance. The SortPooling layer extracts and sorts the vertex features based on the structural roles within the graph\cite{zhang2018end}. Here, the extracted vertex features are the continuous Weisfeiler Lehman (WL) colors. As with the DGCNN model, after we sort all the vertices using the last output layer's output, the $topk$ sorted vertices are chosen to inject the semantic {\tt Nops}.
The transformation to insert $a_t$ into the $topk$ basic blocks is applied to the original graph $G$. We get a new state $G'$ which is the input of the malware detection model $\mathbb{C}$ to get the rewards $r_t \in \mathbb{R}$ for the action.
We obtain the reward for the generated graph $G'$ and store the current state, actions, rewards and next state to the memory buffer $\mathcal{M}$. 
The buffer records past experience denoted as $(s_t, a_t, v_t, r_t, s_{(t+1)}, |\delta_t|)$ with states, actions taken at those states, the rewards and the next state and its absolute temporal difference(TD) error. 
We use the prioritized experience replay technique with memory buffer $\mathcal{M}$ to train the Q network. 
For every $T$ queries, a mini-batch set of samples from the memory buffer are selected with probability $P(i)$ calculated by the normalized TD error. 
The algorithm runs until $niters$ iterations, the inserted features is more than the defined budget $\Delta$, or the generated samples are misclassified as the target label $\widetilde{y}$. Each attack iteratively modifies a malicious CFG until the malicious CFG is misclassified as a benign program or the maximum number of iterations reaches.

\subsection{Other Semantics-preserving Attacks}
In this section, we introduce the proposed other three attacks: (1) Semantics-preserving Random Insertion(SRI) attack using the idea of random insertion; (2) Semantics-preserving Accumulated Insertion(SAI) Attack using the idea of hill-climbing; and (3) Semantics-preserving Gradient based Insertion(SGI) Attack using the idea of gradient-based insertion.



\subsubsection{SRI Attack}
Algorithm~\ref{alg:ri} in the appendix shows the pseudocode of the SRI attack. In each iteration, after randomly choosing $topk$ basic blocks in the CFG $G=<V, E>$, the algorithm picks up one dead instruction in the list of semantic {\tt Nops}. Next, the algorithm inserts the dead instruction into the chosen basic blocks. Here, to limit the numbers of manipulation on the original CFG, difference between the generated CFG and the original CFG should be less than the predefined value of $\Delta$. The algorithm repeats until $niters$ iterations, because the SRI attack may be identified due to the large number of queries on the malware detection model. 

\subsubsection{SAI Attack}\label{sec:sai}
Instead of directly transforming the original binary with the random decision as in Algorithm \ref{alg:ri}, the SAI algorithm follows a hill-climbing approach\cite{Sharif}. The SAI algorithm declines some decisions if the probability of the target class identification decreases.
For each CFG, attacker probes the detection model $\mathbb{C}(\cdot)$ to retrieve the probability $p$ of the target class. Next, in each iteration, attacker randomly chooses $topk$ basic blocks and one dead instruction in a list of semantic {\tt Nops} $NopsList$. Attackers query the detection model using the transformed CFG 
to obtain the changed probability of the target class, $p'$. The transformation is accepted only if the probability increases. Similar to the SRI attack, the SAI attack can also limit the number of effected basic blocks in each iteration and the total number of inquiries. Manipulation on the original CFG should also be less than $\Delta$. However, attackers require the predicted probability of the targeted model to apply the SAI attack.

\subsubsection{SGI Attack}\label{sec:sgi}
The SGI attack solves the constrained optimization problem in Eq.~\ref{equ:opt} with a gradient-descent algorithm. 
To fit the black box setting, we train a substitute model $\mathbb{C'}$ which approximates decision boundaries of the malware detection model $\mathbb{C}$~\cite{Papernot}. We assume that the attackers have some fundamental 
knowledge of the malware detection model including the input and the expected output. The substitute model $\mathbb{C'}$ is trained iteratively with the graphs and the predicted labels.

The substitute model $\mathbb{C'}$ is used when generating adversarial samples. In Algorithm \ref{alg:sg} (in the appendix), we show the pseudocode for generating the adversarial samples. Here, $N$ is the number of basic blocks in the CFG, and $K$ is the number of semantic {\tt Nops}. As shown in Algorithm \ref{alg:sg} when generating adversarial samples, the adversary computes the perturbation, e.g., the signed gradient from the $i_{th}$ iteration, as follows:
\begin{equation}
    g_{i} = sgn(\frac{ \partial J_{\mathbb{C}}(G,y)}{ \partial V}),
\end{equation}
where $V$ is the feature matrix, each row of which describes instructions in a basic block $v_i$ of the graph $G$, and $y$ is the label of the CFG. Attacker heuristically inserts a semantic {\tt Nops} that is closest to the gradient $g_i$ into the corresponding basic block of the CFG. In each iteration, attacker injects the closet semantic {\tt Nops} to the sign gradient descent. Attacker repeats this procedure until a maximum number of iterations $T$.


\section{Assessing the resilience of GNN for malware detection}\label{sec:exp}

We conducted experiments to answer the overarching research question, "\textit{What kind of manipulations can be applied on original CFGs to avoid the malware detection models without changing the programs' behaviors?}". We evaluated the performance of the proposed semantics-preserving attack under various parameters such as the impacts of graph size and the semantic {\tt Nops}. From the experiments, we observed that: 1) the proposed semantics-preserving attack achieves the high evasion rate; 2) it generates small-scale manipulations on original features to succeed attacks; 3) it does not change malicious behaviors while avoiding malware detection model. 


\subsection{Experimental Environment}

To evaluate the performance of the proposed semantics-preserving attacks, we conducted experiments using the malware collected from the VXHeavens Dataset \cite{zhang2013class} and the VirusShare website~\cite{vs}. We used VirusTotal \cite{vt} to label the malware samples and identify the malware families. 
We also used three families of Windows malware samples such as Trojan samples such as Downloader and FakeAV, Virus samples such as HLLC and HLLO, and Backdoor samples such as Simda and Haxdoor.
For benign programs, we installed standard packages on a x86 Windows 10 virtual machine using Ninite and Chocolatey2 package managers and collected the benign binaries generated by those packages. The categories of the installed packages vary including popular applications,
security applications,
developer tools
, and so on. We use a Python framework for analyzing binaries, called Angr\cite{shoshitaishvili2016state}, to extract CFG in those datasets. After extracting CFGs, we transform the basic blocks in CFGs to a directed graph. Because some graphs have millions of basic blocks, it is impossible to train on the entire graph at once due to GPU memory and training time constraints. In our experiments, the CFGs with less than 3,000 basic blocks are used for malware classification. 
Next, we mix benign programs and malware together, and randomly select 20\% samples as the test dataset. 
We trained the model on 70\% samples and validated the model on 10\% samples. Also, to reduce variability on a limited data sample, we trained the evaluation model using 7-fold cross-validation.

All experiments are conducted on Ubuntu 16.04, using Python 3.7 and Tensorflow 2.1 with NVIDIA GTX980 Ti Graphics Processing Unit(GPU). We trained two GNNs as malware detection models over CFG-based features: the basic GCN model; and the DGCNN model. The basic GCN model stacks four graph convolution layers with 128, 64, 32, 16 output channels and use the last graph convolutional layer followed by a fully-connected layer. The DGCNN model\cite{zhang2018end} concatenates four graph convolution layers with 32, 32, 32,
1 output channels, followed by a SortPooling layer to keep top 1000 nodes. Two 1-D convolutional layers with 16 and 32 output channels followed by one dense layer with 128 hidden units are applied to obtain the graph label. We trained two models for 200 epochs with the batch size 100. In Table \ref{tab:detectionfinal}, we show the classification performance of the malware detection models used in the following experiments. The result of other detection models are listed in Table \ref{tab:detection} (in the appendix).
In Table \ref{tab:dataset} (in the appendix), we show the composition of the dataset and the node size distribution for each type of program. The mean value of the Virus CFGs’ node sizes is smaller than the mean value of node sizes of CFGs from other malware families and benign programs, but generally all types contain large and small graphs and are uniformly distributed in the train, validation and test dataset. We generated 28 semantic {\tt Nops} based on rules mentioned in section \ref{sec:nop}. In Table \ref{tab:sno}, we show some examples of semantic {\tt Nops}.

\begin{table}[h]
  \caption{Classification performance}\vspace{-2mm}
  \centering
  \label{tab:detectionfinal}
  \scriptsize
  \begin{tabular}{cccccc}
    \toprule
    \multirow{2}{*}{Model} & \multicolumn{3}{c}{ACC(\%)}& \multirow{2}{*}{FPR(\%)} & \multirow{2}{*}{FNR(\%)}\\
    & Train& Val& Test &&\\
    \midrule
    Basic GCN&97.35&93.61&93.92&5.52&6.62\\
    DGCNN&98.37&93.61&95.77&4.70&3.73\\
  \bottomrule
\end{tabular}
\end{table}

\begin{table}[t]
  \caption{Examples of semantic {\tt Nops}}
\vspace{-2mm}
  \label{tab:sno}
  \centering
  \scriptsize
\begin{tabular}{|ll|}
\hline
    NOP&\\
    PUSH \%rbx& POP \%rbx \\
    NOT  \%rbx& NOT \%rbx \\
    XCHG  \%rax,\%rax& XCHG \%rax,\%rax \\
    ADD  \$5,\%r10& SUB \$5,\%r10\\
\hline
\end{tabular}
\end{table} 

 



\begin{table}[t]
 \caption{Number of attack successes for different families of malware samples}
\vspace{-2mm}
 \centering
  \label{tab:num_attacks}
  \tiny
  \begin{tabular}{ccccccccc}
  \toprule
  \multirow{2}{*}{Attack}&\multicolumn{4}{c}{Basic GCN}&\multicolumn{4}{c}{DGCNN}\\
  &Trojan&Virus&Backdoor&ER(\%)&Trojan&Virus&Backdoor&ER(\%)\\
  \midrule
  IPR&	0&	0&	2& 0.25&	10&	2&	6&2.32\\
Disp-5&6&	1&	2&1.16&30&	3&	15&6.19\\
IPR+Disp-5&7&	2&	2&1.42&	29&	9&	13&6.58\\
  SRI&140&77&156&45.58&216&112&233&72.25\\
  SAI&293&160&300&97.27&288&158&303&96.74\\
  SGI&120&44&153&41.22&239&127&255&83.08\\
  SRL&298&162&311&100.0&297&163&315&100.0\\
  \bottomrule
  \end{tabular}
 \begin{tablenotes}
\item $^{\rm *}$ER=Evasion Rate


\end{tablenotes}
\end{table}

\subsection{Evaluation Results}

To fool the malware detection model, we generated the proposed semantics-preserving attacks by inserting semantic {\tt Nops}. By default, the parameter values for the attacks are configured as follows. Each attack probes the detection model less than 30 iterations. We set 5\% as the maximum injection budget. In the SRI, SAI, and SRL attacks, the maximum effected basic blocks in each iteration is set as 1250. If the number of the nodes in a CFG is less than 1250, one semantic {\tt Nop} is inserted into all basic blocks each step. The norm value of the distance function in the SGI attack is set as 2 to calculate the semantic {\tt Nops} that is close to the direction of the gradient.

We compared our work with the black box algorithm of the binary-diversification attack~\cite{Sharif}. The main concept of the black-box attack is similar to the SAI attack: it queries the detection model after transforming each function, and accepts the modification only if the probability of the benign class increases. This attack implements two categories of detection misleading binary transformations: the in-place randomization (IPR) work of Pappas et al. \cite{pappas2012smashing} and the code displacement (Disp), proposed by Koo and Polychronakis \cite{koo2016juggling}. 
The implementation of the IRP transformations is available at \url{https://github.com/kevinkoo001/ropf}. We evaluated three variants of  the binary-diversification attack~\cite{Sharif}. 
The Disp-5 variant relies on the Disp transformations and increases binaries’ sizes up by 5\%. The last variant, IPR+Disp-5, combines the IPR and Disp transformations.
The target models of the binary-diversification attack are DNN models trained on raw bytes for malware binary detection. To compare it with our method, after transforming the binary, during each iteration, we first extract the CFG of generated binary and the graphed is represented as explained in the Section III-A.  Then we fed the new graph to the GNN detection model to retrieve the probability of the target class. We executed the black box attacks up to 200 iterations and stopped early if the binaries were successfully misclassified, the size of the new CFG is more than 3000, or the binaries’ sizes increase by 5\%. 

For the SGI attack, we trained two substitute models using the validation dataset to approximate the malware detection models\cite{Papernot}. The substitute models have the same structure as the basic GCN model. Both of them are trained for 10 epochs from scratch with CFGs and their corresponding predicted labels. 
After 10 epochs, while the accuracy of the substitute model for the basic GCN model reached by 85.83\% on the test dataset, the substitute model for the DGCNN model showed an accuracy by 81.42\%.

For the SRL attack, we used the RMSProp algorithm with minibatches of size 512.
We trained the model with malicious samples in the training dataset. Each graph can query the detection model for at most 30 steps. We used a replay memory of 3000 most recent queries. The policy during training continuously decayed with $\epsilon$ decaying linearly from 1 to 0.1 over the first 3000 queries, and fixed at 0.1 thereafter. The agent drops actions that returned negative actions with probability 50\% instead of keeping every queries.

In Table \ref{tab:num_attacks}, we show the number of attack successes and the average performance for different families of malware samples against two malware detection models using seven attacks. The binary-diversification attack~\cite{Sharif} showed the evasion rate less than 10\%. We have two observations for the binary-diversification attack.
1) We observed the IPR based transformation is not capable to attack the graph-based models. First, the node features we used is the summation of one-hot encoded opcodes within the basic blocks. Thus, reordering instructions and reassigning registers will not affect the representation of the node features. Second, it only substitutes or reorders the instructions, which does not modify the structure of the generated CFG. The graph-based models, specifically the graph convolutional layers, learn the high-level substructure features by propagating node features to neighboring nodes and the node itself. Thus, it is hard to evade the GNN based detection models without modify the graph structure.
2) The Disp based transformation create certain jump basic blocks and dead code basic blocks. So the structure of the generated CFG will be changed. Thus, it performs better compared with IPR based transformation. However, the jump basic blocks only contain one opcode and the dead code basic blocks will never be reached, so they have small effect when calculating the graph features.

The SRI attack showed the evasion rate by 45.58\% on the basic GCN model when inserting instruction by 0.96\% on average, and by 72.25\% on the DGCNN model when 1.28\% of features changed on average. The SGI attack showed the evasion rate by 41.22\% on the basic GCN model and by 83.08\% on the DGCNN model. Overall, we observed that SAI and SRL attacks have shown a good performance in general. The SAI attack showed the evasion rate by 97.27\% on the basic GCN model with 0.54\% injected features and 96.74\% on the DGCNN model with 0.62\% injected features. 
In practice, SRI attacks took more time when generating an sample compared with SAI attack on average. This is because the SAI attack drops some transformation and causes the misclassification from the detection model within fewer iterations. Also, the SGI attack costs longer time because of the gradient calculation. 
The SRL attack against both models showed the evasion rate by 100\% with 0.17\% injected features on the basic GCN model and 0.23\% injected features on the DGCNN model. 
SRL altered CFGs by inserting semantic {\tt Nops} into the corresponding basic blocks. 

To explain why the SRL attack achieves a high attack rate, we analyzed the frequency of the inserted features. The semantic {\tt Nops} used in our experiments contains 28 opcodes. In Figure \ref{fig:freq1} (in the appendix), we show the occurrence frequency of those opcodes in our dataset. Even though some opcodes such as $CMOVA$, $CMOVL$, and $CMOVNS$ are not commonly seen in the whole datasets, there is a significant discrepancy between benign and malicious programs. For example, the frequency of appearance of $CMOVA$ in benign programs is $6.44e^{-5}$, which is different from that in malicious programs $2.12e^{-5}$. In Figure \ref{fig:freq2}  (in the appendix) and Figure \ref{fig:freq3} (in the appendix), we show the opcodes chosen by different attacks for the DGCNN model and the basic GCN model. We run four attacks on each detection model 10 times and calculate the inserted opcodes of the adversarial samples. We observed the SRL attack highly relies on certain opcodes that have a substantial difference, such as $CMOVG$, $CMOVA$, and $CMOVS$, to evade detection. The SRL attack seems to strategically perform less frequently used opcodes, which implies that the detection models learn those differences as graph features to detect malware and the SRL attack effectively finds the blind spot of the detection models.



\subsubsection{Impact of Various Attack Parameters}

\begin{figure*}[t]
     \centering
     \begin{subfigure}[b]{0.3\linewidth}
         \centering
         \includegraphics[width=\linewidth]{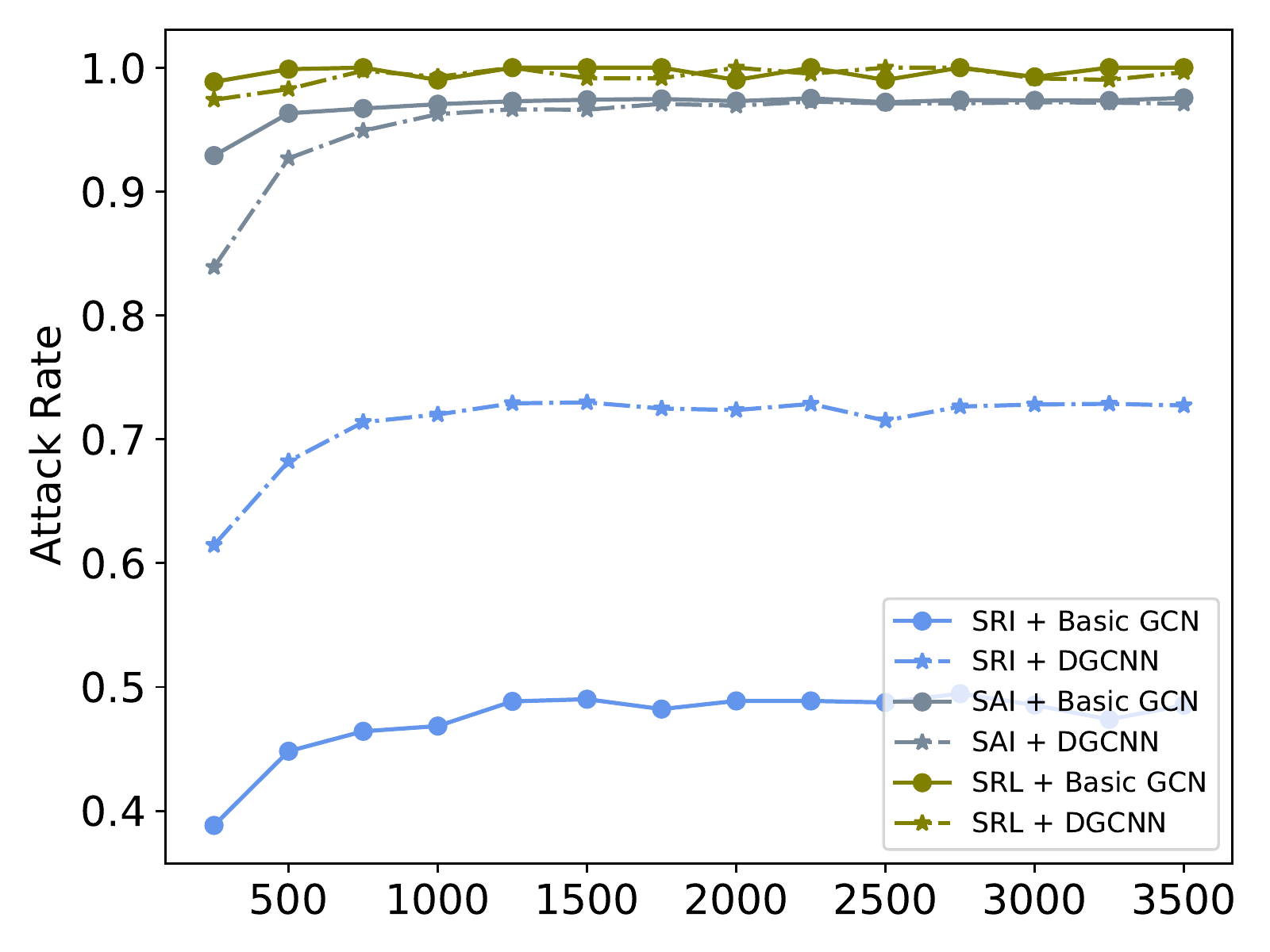}
         \caption{Impact of the basic blocks}
         \label{fig:topk}
     \end{subfigure}
     \hfill
     \begin{subfigure}[b]{0.3\linewidth}
         \centering
         \includegraphics[width=\linewidth]{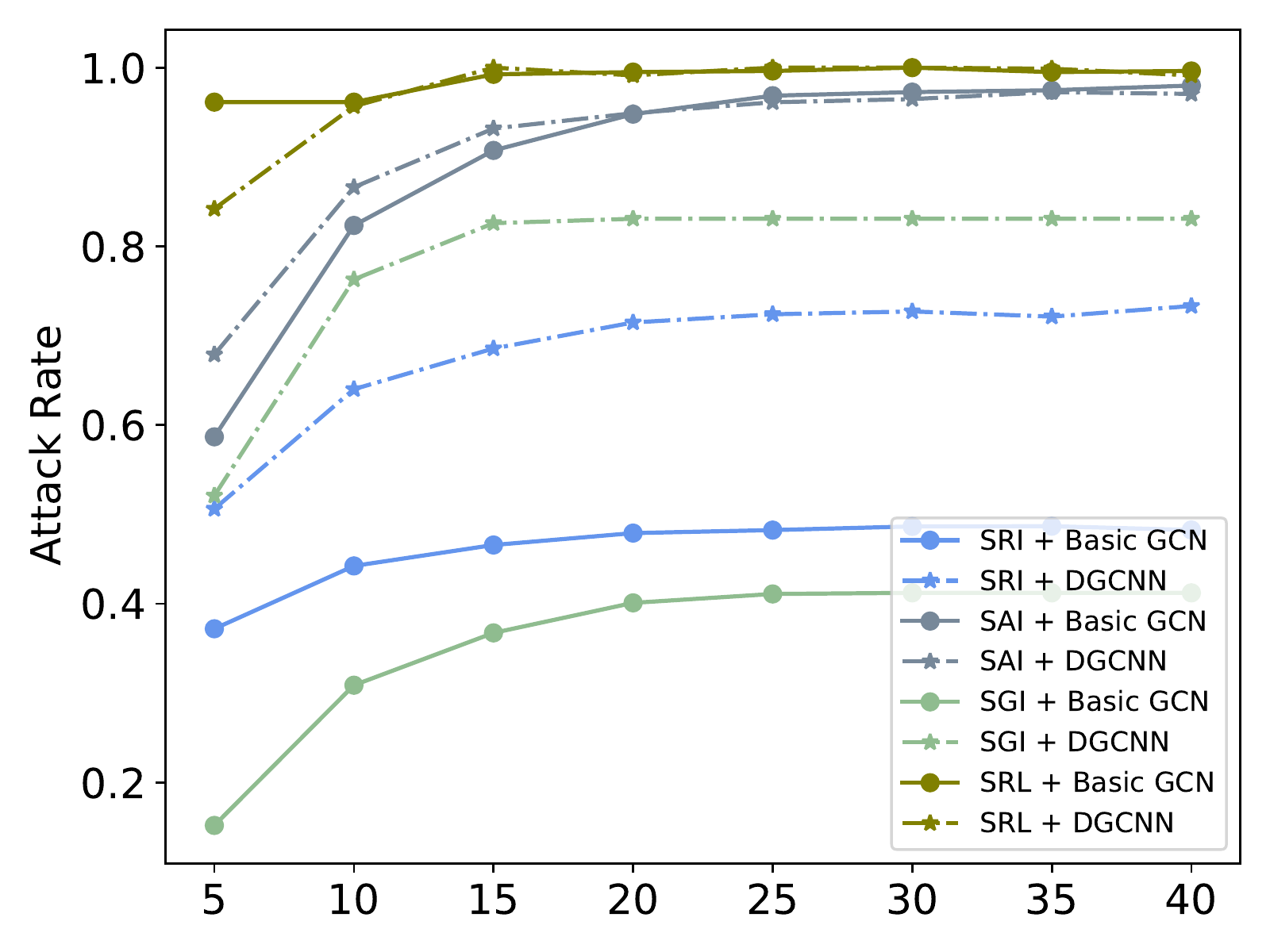}
         \caption{Impact of iteration}
         \label{fig:iteration}
     \end{subfigure}
     \hfill
     \begin{subfigure}[b]{0.3\linewidth}
         \centering
         \includegraphics[width=\linewidth]{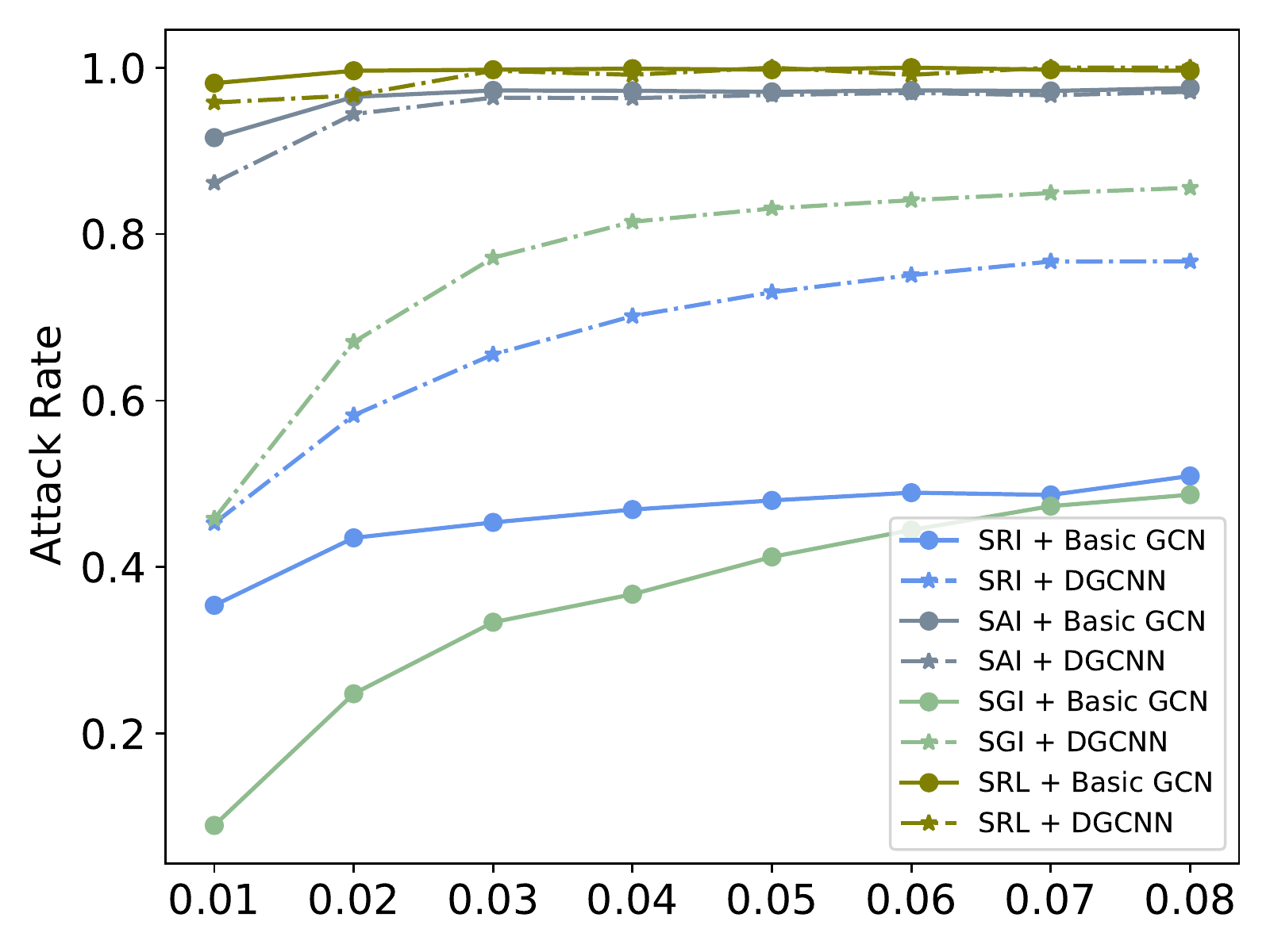}
         \caption{Impact of insertion budget}
         \label{fig:injection}
     \end{subfigure}
        \caption{Impact of various attack parameters}
        \vspace{-4mm}
\end{figure*}

In this section, we discuss the impact of various attack parameters including the number of effected basic blocks, maximum iterations, and insertion budget.

We suggest choosing the value of $topk$ based on the node size of the CFGs. If the value of $topk$ is too large, it causes large-scale manipulation on original features and can affect the functionality of the program. On the other hand, if it is too small, it requires more iterations to evade the detection models. In practice, we can determine the value of $topk$ from the empirical experiments which compare the performance under some reasonable values.
In Figure \ref{fig:topk}, we shows evasion rates of SRI, SAI and SRL attacks under different numbers of effected basic blocks in each iteration. Here, the number of iterations and the insertion budget are set as 30 and 5\% respectively. We immediately observed that more effected basic blocks do not significantly improve the evasion rate after the number of effected basic blocks reaches 1250. That is because when the number of the effected basic blocks is more than the node size of the CFG, one dead instruction is inserted to all basic blocks.
With more effected basic blocks, we observed that the evasion rate tends to depend on other factors such as the number of iterations and the insertion budget as shown in Figure \ref{fig:iteration} and Figure \ref{fig:injection}.



In Figure \ref{fig:iteration} and Figure \ref{fig:injection}, we show how the evasion rates on two detection models vary under different maximum iterations and insertion budget. Generally, we observed that attacks with more iterations and higher budget are more successful. Four attacks showed a good performance at about 30 iterations. Also, most attacks showed a good performance when insertion budgets reached 5\%. As one would expect, more iterations and insertion budget change more features when generating an adversarial sample. For example, for SGI attack against the DGCNN model, the number of changed features increased from 0.57\% to 1.22\% with the increase of the insertion budget on average. 

Even though the insertion budget is small, e.g. 1\%, we observed that the SRL attack showed the high evasion rate close to 100\%. However, let us note that the evasion rate is highly dependent on the initial value of the action-value function. If the evasion rate before training the Q function is less than 40\%, most queries in the earlier training process will get negative rewards, and consequentially need more time when training the Q function for a good result. Moreover, smaller iterations and insertion budget restrict the length of action sequences, which may reduce the number of queries with positive rewards. We also observed that although a smaller number of iterations and insertion budget obtained a good performance, it might take longer time to train the Q function because most queries in the beginning of the training process return negative rewards.

\subsubsection{Impact of Graph Size}

We measured the evasion rates of four attacks under various size of graph using the test dataset, which is separated into four groups according to the the number of the nodes in a graph. As shown in Table~\ref{tab:size}, we consider the first quartile, the median, and the third quartile of the graph size to group the test dataset. For the SGI attack, larger graphs significantly showed higher evasion rates. However, from the other attacks, we did not observe any meaningful relationship between the graph size and the evasion rate. This observation indicates that the evasion rate is not highly dependent on the graph size.



\subsubsection{Impact of Dead Instructions}
To evaluate the impact of dead instructions, we investigated the performance of a single dead instruction with different attacks on two detection models. In Table \ref{tab:dead}, we show impact of some dead instructions with different semantic $Nop$s.
First, we observe that some opcodes, such as $CMOVA$, $CMOVG$, $CMOVL$ and $CMOVS$, are useful for evading the detection models. This is because the number of those opcodes in the benign and malicious programs is different and the detection models learned this discrepancy.
Second, with some dead instructions, such as $NOT, NOT$, three attacks on different models also showed divergent evasion rates. The key factor that leads to this different impact is two models have different decision boundaries and the DGCNN model holds more fine-grained information collected from the all graph convolution layers which lead it easier to be evaded by inserting simple instructions.
Third, we observed that sometimes the SAI attacks can not show the same performance as the SRI attacks. The SAI attacks reject some insertions and consequentially reduce the action sequences. This strategy achieves a high evasion rate when we do not know which instructions are useful. However, it is hard for reduced action lists to evade the malware detection models. 
Also, some dead instructions may not cause the misclassification. Even using those instructions as candidates, the SAI and SRL attacks can reject those ‘useless’ instructions or locate a better combination to deceive the detection model.

\begin{table}[h]
 \caption{Impact of graph size(\%)}\vspace{-2mm}
 \centering
  \label{tab:size}
  \small\addtolength{\tabcolsep}{-0.2pt}
  \scriptsize
  \begin{tabular}{ccccccccccc}
  \toprule
  \multirow{2}{*}{Attack}&\multicolumn{4}{c}{Basic GCN}&\multicolumn{4}{c}{DGCNN}\\
  &25\%&50\%&75\%&100\%&25\%&50\%&75\%&100\%\\
  \midrule
  SRI&44.44&51.44&47.44&53.64& 70.81&71.29&77.94&79.47 \\
  SAI&98.06&98.55&98.97&94.79& 94.73&95.83&97.43&95.78\\
  SGI&10.14&17.30&54.59&86.97& 54.02&82.87&84.10&88.94\\
  SRL&100.0&100.0&100.0&100.0& 100.0&100.0&100.0&100.0\\
  \bottomrule
  \end{tabular}
\end{table}

\begin{table}[h]
  \caption{Impact of dead instructions(\%)}\vspace{-2mm}
 \centering
  \label{tab:dead}
  \tiny
  \begin{tabular}{cccccccc}
  \toprule
   \multirow{2}{*}{ID}& \multirow{2}{*}{Semantic Nops} &\multicolumn{3}{c}{Basic GCN}&\multicolumn{3}{c}{DGCNN}\\
   & &SRI & SAI& SGI  & SRI & SAI& SGI \\
    \midrule
    1&NOP&28.51& 38.89& 32.12&95.55&80.49&95.30\\
    2&SUB&0&0 &0&0.24&0.12 &0.24\\
    3&ADD& 54.91&55.29 &57.53& 67.16& 47.77&67.90\\
    4&LEA&14.57& 13.32&14.44& 2.22 & 4.07&2.46\\
    5&MOV&7.97&7.97 &8.09&81.85 &41.23&84.69\\
    6&XCHG&0&0& 0 &4.19&2.09&4.07\\
    7&CMOVO&0& 0&0 & 77.16&22.96&80.98\\
    8&CMOVP&0&0 & 0&86.54&90.74&91.23\\
    9&CMOVA & 98.14&99.25&99.25&100&100&100 \\
    10&CMOVG&100& 100&100 &99.87&96.54&99.87\\
    11&CMOVS&100&100 &100&70.00&63.70&68.76 \\
    12&CMOVL&99.25& 100&99.25 &77.65&33.95 &81.72\\
    13&CMOVNS&0&0& 0&0&0 &0\\
    14&CMOVNP& 0&0 &0&91.48&86.67&84.81 \\
    15&CMOVNO & 0&0 &0&94.19& 66.41&93.33\\
    16&ADD,SUB&0& 0& 0&0.24 &0.37&0.24 \\
    17&SUB,ADD&0& 0&0&1.23&0.24 &0.24\\
    18&NEG,NEG&87.17& 71.98&86.05&99.75 &97.03&99.50\\
    19&NOT,NOT&0& 0&0& 53.33&42.71&49.75\\
    20&PUSH,POP&2.24& 2.36&2.24&53.08&30.24&52.83 \\
    21&PUSHF,POPF&0& 0&0 & 4.32&1.11 &8.76\\
    22&XCHG,XCHG& 0&0 &0&5.18&3.08 &9.38\\
    23&BSWAP,BSWAP&0&0&0  &10.49&3.70 &10.24\\
    24&PUSH,NOT,POP&0& 0& 0&47.40&35.92&43.20 \\
    25&XOR,XOR,XOR&0&0 &0 &96.79&40.37&90.86 \\
    26&MOV,ADD,MOV & 3.73&3.73&3.73&81.72 &64.56&82.09 \\
    27&INC, PUSH, DEC, DEC &4.23& 2.11&4.23&46.54&13.95&47.28 \\
    28&MOV,CMP,SETG,MOVZX,MOV,MOV&98.38&98.38 &98.38 &75.30&74.07&74.56 \\
    
  \bottomrule
\end{tabular}
\end{table}

\begin{table}[t]
  \caption{Classification performance after retraining the model}\vspace{-2mm}
  \centering
  \label{tab:redetection}
  \scriptsize
  \begin{tabular}{cccccc}
    \toprule
    \multirow{2}{*}{Model} & \multicolumn{3}{c}{ACC}& \multirow{2}{*}{FPR} & \multirow{2}{*}{FNR}\\
    & Train& Val& Test &&\\
    \midrule
    Basic GCN&98.96\%&93.33\%&94.70\%&4.94\%&5.66\%\\
    DGCNN&98.71\%&93.61\%&95.23\%&3.41\%&6.14\%\\
  \bottomrule
\end{tabular}
\vspace{-3mm}
\end{table}

\begin{table}[t]
 \caption{Impact of a potential mitigation method based on retraining}\vspace{-2mm}
 \centering
  \label{tab:defense}
  \scriptsize
  \begin{tabular}{ccccc}
  \toprule
  \multirow{2}{*}{Attack}&\multicolumn{2}{c}{Basic GCN}&\multicolumn{2}{c}{DGCNN}\\
  &ER(\%)&FG(\%) &ER(\%)&FG(\%)\\
  \midrule
  SRI&1.05&0.91&2.02&1.01\\
  SAI&13.24&0.94&16.96&0.94\\
  SGI&0.24&2.94&0.73&1.03\\
  SRL&27.22&1.20&85.74&0.83\\
  \bottomrule
  \end{tabular}
  \vspace{-4mm}
\end{table}

\subsection{Potential Mitigation}\label{sec:disc}
In this section, we briefly discuss a potential mitigation method to make malware detection robust against the proposed semantic-preserving attacks. To defend against the proposed semantic-preserving attacks, we retrained the detection model using the original training dataset together with the adversarial samples generated by the attacks. For each attack, we randomly selected 500 adversarial samples and added them into the training dataset. As shown in Table \ref{tab:redetection}, detection models showed the same accuracy as the original one after 200 epochs. For SRI, SAI and SGI attacks, the evasion rate of the retrained model significantly decreased as shown in Table \ref{tab:defense}. Let us note that even though such a defense method is effective, the cost of retraining the detection model is large if the training set is large. Also, to decrease the susceptibility against those attacks, we recommend to design a malware detection system which combines multiple detection algorithms. 






\subsection{Discussion}
The proposed SRL attack generates a new CFG by inserting semantic {\tt Nops} into the  CFG,  so it does not generate actual malware.  In practice,  the inserted feature will increase the size of each basic block which also changes the position of other code and data.  Generally speaking,  there are three methods to modify the original binary based on the generated adversarial CFG  \cite{hai2015obfuscation}: (1) Source code level\cite{chen2016remix}: the compiler is modified to insert enough space and generate certain types of instructions for the basic blocks;  (2) Intermediate code level \cite{crane2015readactor}: the target, e.g. jump table, is modified using intermediate language such as  LLVM  to automatically adjust the addresses; (3) Machine code level \cite{williams2016shuffler,chen2017you}: the last instruction of each basic blocks can be replaced as a jump instruction to jump to a trampoline function which helps to calculate the real address.  It  should  be  noticed  that  after  those  methods are applied, the functionality of the original malware will not be changed.

In our current experiment, the SRL attack is adaptive for different detection models. The generated perturbations are different for the basic GCN model and the DGCNN model. This indicates the SRL attack certainly could fool other graph based detection models. To further test the hypothesis, we trained another graph-based detection model, called the GAT (graph attention networks) model, and applied four attacks on it. The key difference between the GAT model and the GCN model is how they generalize the information of the nodes and their neighborhoods. The GCN model calculates the normalized sum of the node features of neighbors. For the GAT model, instead of applying the graph convolution operation, it performs the self-attention mechanism on the nodes by computing the attention coefficients of each node over their neighborhoods’ features. The GAT model stacks four graph attentional layers to extract the graph features. After adding one classification layer to summarize the substructure features, the GAT model makes predictions for the CFGs. The model achieved an accuracy of 88.62\% on the test dataset. For the SGI attack, the structure of the substitute model of the GAT model is the same as the basic GCN model and it can obtain an accuracy of 82.32\% on the test dataset. Also, the SRI, SAI, SGI, and SRL attacks achieved 18.79\%, 81.76\%, 84.43\%, and 98.03\% respectively.

\section{Related Work}\label{sec:related}
In this section, we introduce research on attacking neural networks and malware detection models.

\subsection{Attacks on Neural networks}
Adversarial samples are generated by adding imperceptible perturbations to original samples to deceive deep learning algorithms. Attacks can be grouped into three scenarios based on their knowledge: white-box attack, black-box attack, and semi-white (gray) box attack.

In the white-box scenario, attacks have access to the architecture and parameters of the neural networks. 
The FGSM attack\cite{goodfellow2014explaining} is a fast method to generate adversarial samples.
Carlini and Wagner \cite{carlini2017towards} proposed gradient-based attacks to generate adversarial samples by calculating one back-propagation step. 
In a black-box attack setting, the architecture parameters of the neural network is unavailable to attackers.
Attackers only have the query access to generate adversarial samples.
Papernot et al. \cite{Papernot} design a substitute neural network to fit the black-box neural network and then generated adversarial examples according to the substitute neural network. This method assumes the attackers can only obtain the label information from the target neural network. Zeroth order optimization based black box attack has a different assumption that the attackers have access to the prediction confidence (score) from the target neural networks\cite{chen2017zoo}. 
In the grey-box setting, attackers have access to the the structure of the target model. Generative Adversarial Network(GAN) is introduced to generate adversarial examples directly from the generative network\cite{xiao2018generating}.


\subsection{Attacks on Malware Detection}
Adversarial malware generation contrasts with previous applications of adversarial sample generation in computer vision because most features in malware detection involve in discrete data, which means the gradients to train the generator are zero almost everywhere. 
There are two constraints for adversarial malware generation: 1) large-scale manipulations on original features may change the program's functionality; 2) generated samples should not remove original features\cite{Wang2017}. 
Multiple attacks are introduced to evade deep learning models. Some of them modify original malware, for example, add benign code to malware, to mimic benign program\cite{laskov2014practical,Kreuk,Sharif}.
Anderson et al.\cite{anderson2018learning} modify PE header metadata, Section metadata, Import and Export Table metadata, etc. by  reinforcement learning to evade static PE machine learning malware models.
Park et al.\cite{Park2019} generate the executable adversarial malware examples by inserting semantic {\tt Nops} to evade CNN based detection models. 
Other attacks focus on the gradient of the detection model or the substitute model to tweak some features of the target model, e.g. adding new API calls \cite{Al-Dujaili2018}.
The evolutionary computation techniques are used to develop new variants of mobile malware\cite{Sen2018}. 

Although researchers have been developing graph based neural network models for malware detection, no systematic study on whether such models can be attacked. 
Unlike to image data, it is harder to generate adversarial samples on graph-based data for two reasons: 1) the graph structure is discrete so we cannot use infinitesimal small perturbation, and 2) large graphs can not easily be verified visually. 
The attacks on graph based model can be grouped into two categories: gradient-based attack and non-gradient-based attack\cite{sun2020adversarial}. Gradient-based attacks retrieve or estimate the gradient information of the detection model to modify the original samples.
Chen et al.\cite{chen2018fast} introduced a network embedding attack that uses the gradient information of the adjacency matrix to iteratively add or delete some key links.
Zügner et al.\cite{zugner2018adversarial} first proposed a method $Nettack$ to perturb the graph data to perform poisoning attack on GCN model. 
Non-gradient-based attacks solve the graph based optimization problem without using the gradient of the detection model.
Dai et al.\cite{dai2018adversarial} proposed a reinforcement learning based attack method  RL-S2V that learns to modify the graph structure by sequentially adding or dropping edges from the graph. 
Wang et al.\cite{wang2018attack} developed two algorithms, Greedy and Greedy-GAN, to attack GCN models by adding fake nodes into the original graph.
Wang et al.\cite{Wang00553} formulated a graph-based optimization problem to manipulate edges and solved the problem using projected gradient descent method.
Sun et al.\cite{sun2019node} extend the idea of RL-S2V to sequentially add fake nodes, introduce fake links, and modify the labels of fake nodes. 

Differently from some of the prior works, e.g. \cite{Sharif,Al-Dujaili2018}, our target models are graph based deep learning models from CFGs. The generated samples should not change the program's functionality. So we can not arbitrarily add or remove the edges and nodes features of the CFGs.
In our paper, we focus on generate adversarial CFGs by iteratively insert semantic {\tt Nops} into original graphs. 
Perhaps most closely related to our work on evading CFG based deep learning model\cite{Abusnaina}. First, instead of using graph neural networks,  
they extracted some indicators such as closeness centrality, density, and betweenness centrality, to represent the CFG and constructed a detection model. Because those features are continuous value, they can directly apply methods of white box attack to generate adversarial samples. However, those generated adversarial samples can not be reverted to CFG and might change the structure, features, and consequently, the programs' behaviors. The graph embedding and augmentation method they proposed reply on expert experience on modify the CFG and doesn't consider the guidance of the malware detection.

\section{Conclusion}
\label{sec:conc}

In this work, we propose a reinforcement learning based 
semantics-preserving 
attack against black-box GNNs. 
The key factor of adversarial
malware generation via semantic {\tt Nops}
insertion is to select the
appropriate semantic {\tt Nops} and their corresponding basic blocks.
The proposed attack uses reinforcement learning
 to automatically make these ``how to select" decisions.
To evaluate the attack, we have 
  trained two kinds of GNNs  with three types 
  of Windows malware samples and various benign Windows programs. 
The evaluation results shown that 
the proposed attack can achieve a significantly higher 
evasion rate than three baseline attacks, namely   
the semantics-preserving random instruction insertion attack,  
the semantics-preserving accumulative instruction insertion attack, 
and the semantics-preserving gradient-based 
instruction insertion attack.


Our work focuses on attacks targeting GNNs for malware detection using CFGs as input. 
We believe the method can be applied to other ML models based on sequential data. 
For example, prior work studied the effectiveness of instruction sequences and API Call sequences as 
features for malware detection.
By inserting semantic {\tt Nops} into the instruction sequences, one can 
potentially mislead RNN-based detection models. 
Another potential extension for the SRL attack is to leverage other code obfuscation techniques 
such as instruction substitution and code transposition.



\bibliographystyle{IEEEtran}
\bibliography{bibliography}

\clearpage
\newpage

\appendices
\section{Algorithm of the SRI Attack}
Algorithm \ref{alg:ri} shows the pseudocode of the SRI attack.
\begin{algorithm}[h]
\begin{algorithmic}[1]
\begin{scriptsize}
\renewcommand{\algorithmicrequire}{\textbf{Input:}}
 \renewcommand{\algorithmicensure}{\textbf{Output:}}
 \REQUIRE $\mathbb{C}(\cdot)$,  $G=<V, E>$ , $\widetilde{y}$, $NopsList$, $topk$, $niters$, $\Delta$
 \ENSURE $\widetilde{G}=<\widetilde{V}, E>$
 \STATE  $i \leftarrow  0$ \;
 \STATE  $\widetilde{G} \leftarrow G$ \;
 \WHILE{$\argmax(\mathbb{C}(\widetilde{G})) \neq \widetilde{y}$ and $i < niters$}
  \STATE  $node\_mask \leftarrow RandomBasicBlock(\widetilde{V}, topk)$ \;
  \STATE  $X \leftarrow RandomInstruction(NopsList)$\;
  \STATE  $\widetilde{V} \leftarrow V * node\_mask + (V+X) * node\_mask$\;
  \STATE  $G'\leftarrow<\widetilde{V}, E>$\;
  \IF{$Diff(G', G) <= \Delta$}
  \STATE  $\widetilde{G}\leftarrow G'$\;
  \ENDIF
  \STATE  $i\leftarrow i+1$
 \ENDWHILE
\end{scriptsize}
 \end{algorithmic}
 \caption{SRI Attack against Malware Detection}
 \label{alg:ri}
\end{algorithm}

\section{Algorithm of the SGI Attack}
Algorithm \ref{alg:sg} shows the pseudocode of the SGI attack.

\begin{algorithm}[h]
\begin{algorithmic}[1]
\begin{scriptsize}
\renewcommand{\algorithmicrequire}{\textbf{Input:}}
 \renewcommand{\algorithmicensure}{\textbf{Output:}}
 \REQUIRE $\mathbb{C}(\cdot)$,  $G=<V, E>$, $y$, $\widetilde{y}$, $NopsList$, $niters$, $\Delta$, $N$, $K$
 \ENSURE $\widetilde{G}=<\widetilde{V}, E>$
 \STATE $i \leftarrow  0$ \;
 \STATE  $\widetilde{G} \leftarrow G$ \;
 \STATE  
 \WHILE{$argmax(\mathbb{C}(\widetilde{G})) \neq \widetilde{y}$ and $i < niters$}
  \STATE $g_{i}\leftarrow sgn(\frac{ \partial J_{\mathbb{C}}(G,y)}{ \partial V})$\;
  \FOR{$i \leftarrow 1$ to $N$} 
      \FOR{$j \leftarrow 1$ to $K$}
        \STATE $d_j \leftarrow \left \|  g_i - NopsList_j\right \|_p$\;
      \ENDFOR
      \STATE $X_i \leftarrow NopsList_{argmin(d_j)}$\;
  \ENDFOR
  \STATE $\widetilde{V} \leftarrow V + X$\;
  \STATE $G'\leftarrow<\widetilde{V}, E>$\;
  \IF{$Diff(G', G) <= \Delta$}
    \STATE $\widetilde{G}\leftarrow G'$\;
  \ENDIF
  \STATE $i\leftarrow i+1$
 \ENDWHILE
\end{scriptsize}
 \end{algorithmic}
 \caption{SGI attack against malware detection}
 \label{alg:sg}
\end{algorithm}

\section{Malware Classification Results}
Table \ref{tab:detection} shows the performance of malware classification.
\begin{table}[h]
  \caption{Classification performance}\vspace{-2mm}
  \centering
  \label{tab:detection}
  \scriptsize
  \begin{tabular}{ccccccc}
    \toprule
    &\multirow{2}{*}{Model} & \multicolumn{3}{c}{ACC(\%)}& \multirow{2}{*}{FPR(\%)} & \multirow{2}{*}{FNR(\%)}\\
    && Train& Val& Test &&\\
    \midrule
    1&Basic GCN&98.58&93.81&93.75&7.37&6.36\\
    &DGCNN&94.57&93.50&94.12&4.54&7.25\\
    
    2&Basic GCN&96.12&92.62&92.81&7.00&7.37\\
    &DGCNN&95.01&93.25&94.00&4.29&7.76\\
    
    3&Basic GCN&96.44&91.37&93.06&8.47&5.34\\
    &DGCNN&95.17&92.62&94.31&5.77&5.59\\
    
    4&Basic GCN&96.44&92.18&92.12&8.23&7.37\\
    &DGCNN&93.64&91.75&93.06&5.03&8.90\\
    
    5&Basic GCN&94.66&89.75&91.81&9.95&6.36\\
    &DGCNN&93.55&92.37&92.81&5.28&9.16\\
    
    6&Basic GCN&99.00&92.12&93.87&4.43&7.88\\
    &DGCNN&91.03&89.00&91.50&8.59&8.39\\
    
    7&Basic GCN&97.35&93.61&93.92&5.52&6.62\\
    &DGCNN&98.37&93.61&95.77&4.70&3.73\\
  \bottomrule
\end{tabular}
\end{table}

\section{Data distribution for each type of program}

Table \ref{tab:dataset} shows the data distribution of the whole dataset.
\begin{table}[h]
  \caption{Data distribution for each type of program}
  \centering
  \label{tab:dataset}
  \scriptsize
  \begin{tabular}{ccccccc}
    \toprule
    \multirow{2}{*}{Dataset} &\multirow{2}{*}{Node size}&\multirow{2}{*}{Benign}&\multicolumn{3}{c}{Malware}&\multirow{2}{*}{All}\\
    & &&Trojan&Virus&Backdoor&\\
    \midrule
    Train&count&2822&1165&531&1082&5600\\
&mean&1148&1111&671&1318&1128\\
&std&948&821&771&938&920\\
&min&14&13&22&22&13\\
&max&3237&3205&3175&3210&3237\\
Val&count&392&165&80&163&800\\
&mean&1133&1063&666&1249&1096\\
&std&974&792&698&916&914\\
&min&28&35&24&58&24\\
&max&3240&3145&2428&3192&3240\\
Test&count&786&318&171&325&1600\\
&mean&1184&1132&747&1280&1147\\
&std&967&817&855&949&934\\
&min&21&28&22&42&21\\
&max&3238&3181&3200&3211&3238\\

  \bottomrule
\end{tabular}
\end{table}

\section{Performance of the SRL attack}
\begin{figure}[h]
     \centering
     \begin{subfigure}[b]{0.75\linewidth}
         \includegraphics[width=1.\linewidth]{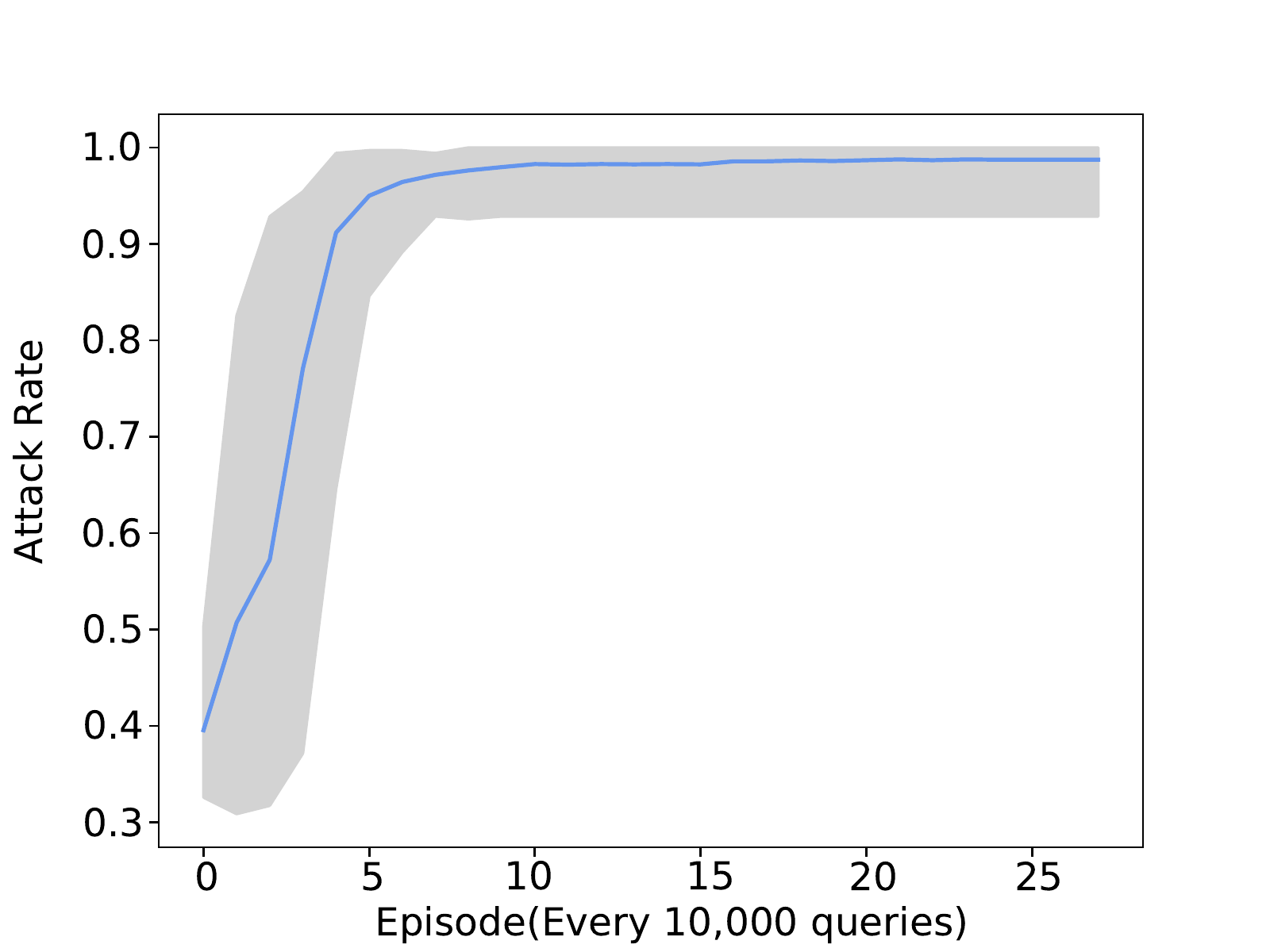}
         \caption{On the basic GCN model}
         \label{fig:srl_gcn}
     \end{subfigure}
     \hfill
     \begin{subfigure}[b]{0.75\linewidth}
         \includegraphics[width=1.\linewidth]{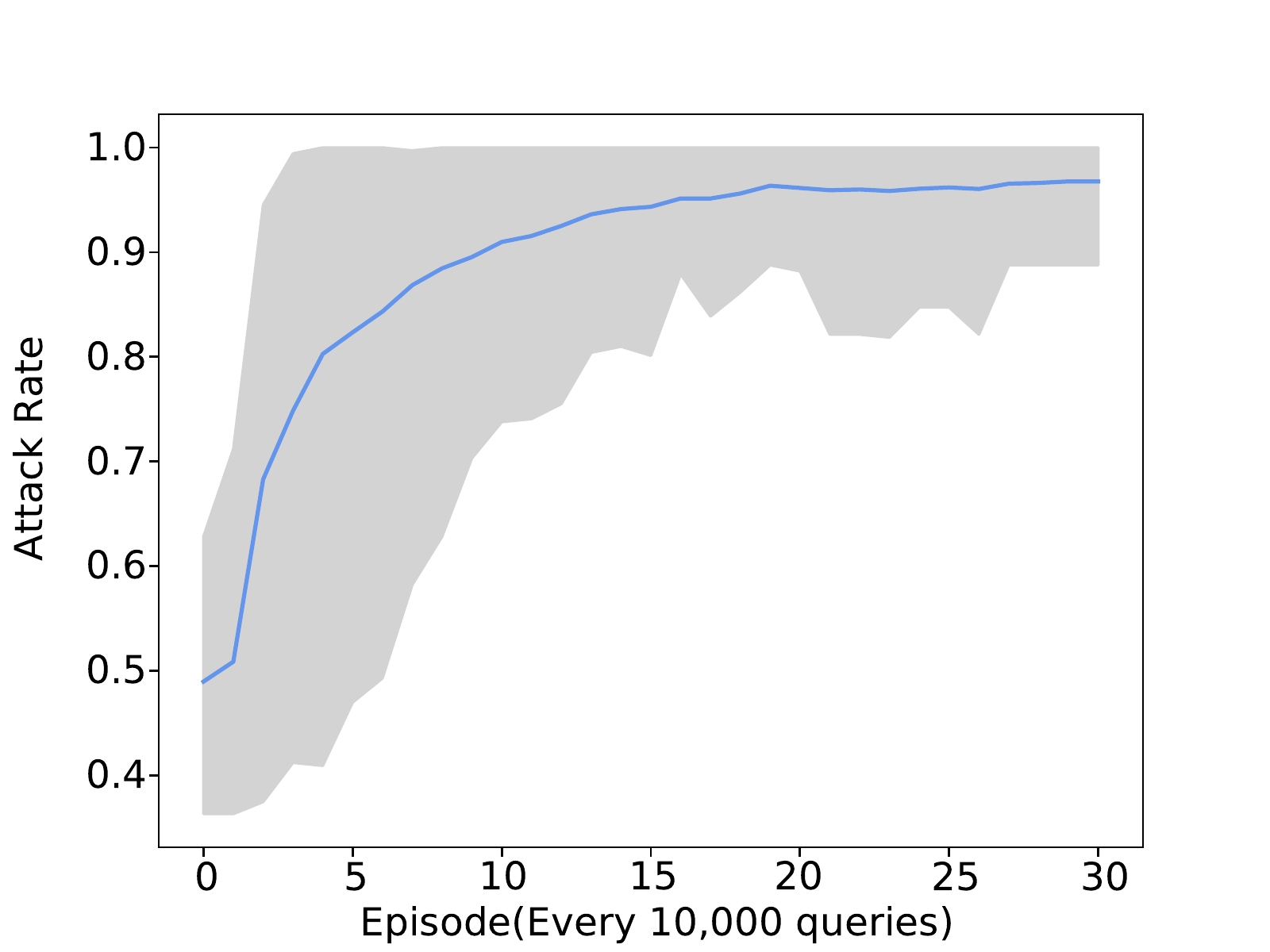}
         \caption{On the DGCNN model}
         \label{fig:srl_gnn}
     \end{subfigure}
     
        \caption{Performance of the SRL attack}
        \label{fig:srlperfo}
\end{figure}

In Figure \ref{fig:srl_gcn} and Figure \ref{fig:srl_gnn}, we show the performance of the SRL attack against the basic GCN model and the DGCNN model. We run the attacks against two models 10 times and periodically compute the evasion rate on the validation dataset every 10,000 queries during training. Here, while the bold lines shows average values over 10 independent learning trials, the shaded area show the maximum and minimum values from 10 independent learning trials. We observed that evasion rate against the basic GCN model converged smoother and faster than evasion rate against the DGCNN model. 

\section{Injected Features}
\begin{figure}[h]
     \centering
     \begin{subfigure}[b]{0.8\linewidth}
         \centering
         \includegraphics[width=1.\linewidth]{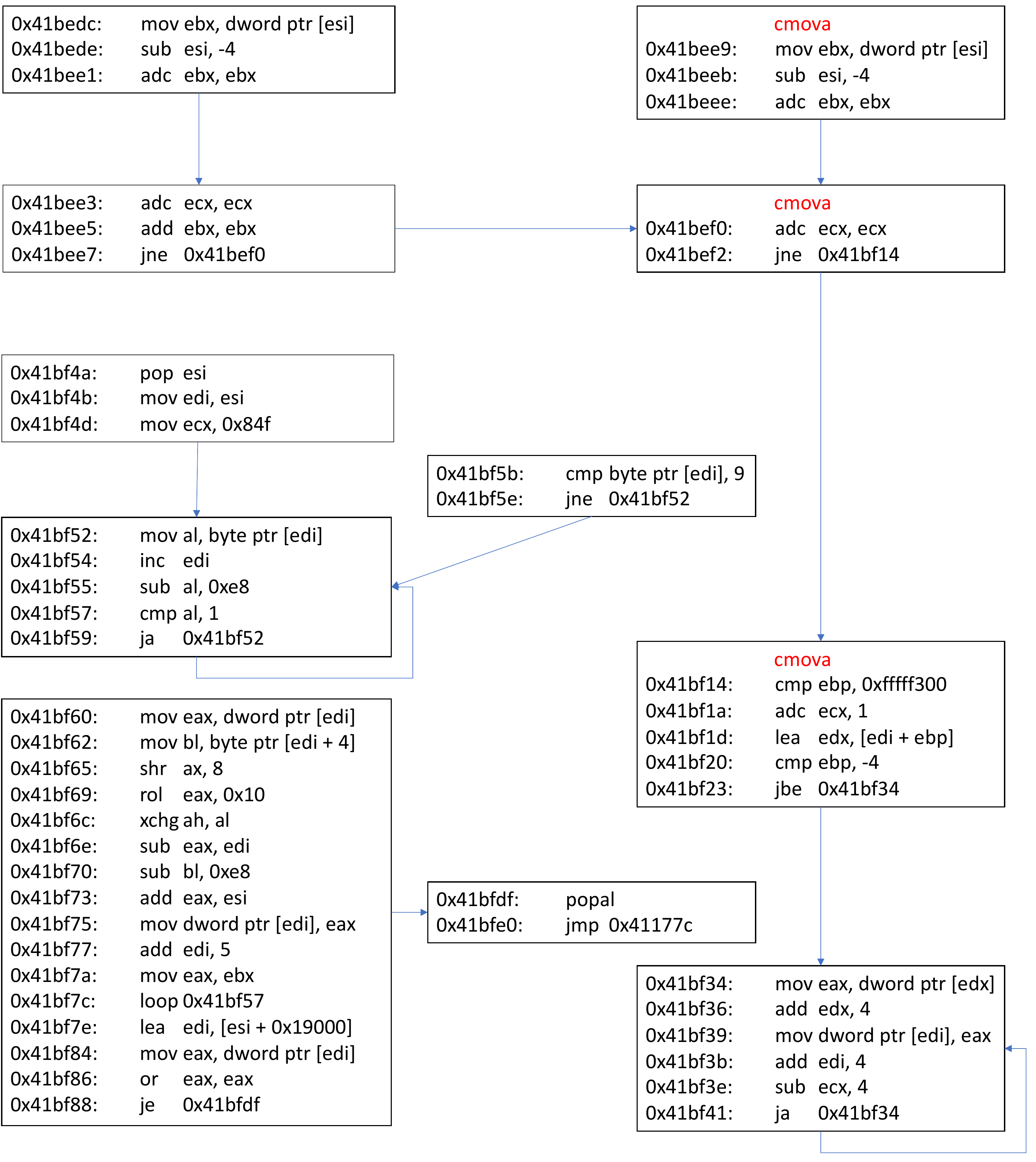}
         \caption{For the basic GCN model  }
         \label{fig:sample_gcn}
     \end{subfigure}
     \hfill
     \begin{subfigure}[b]{0.8\linewidth}
         \centering
         \includegraphics[width=1.\linewidth]{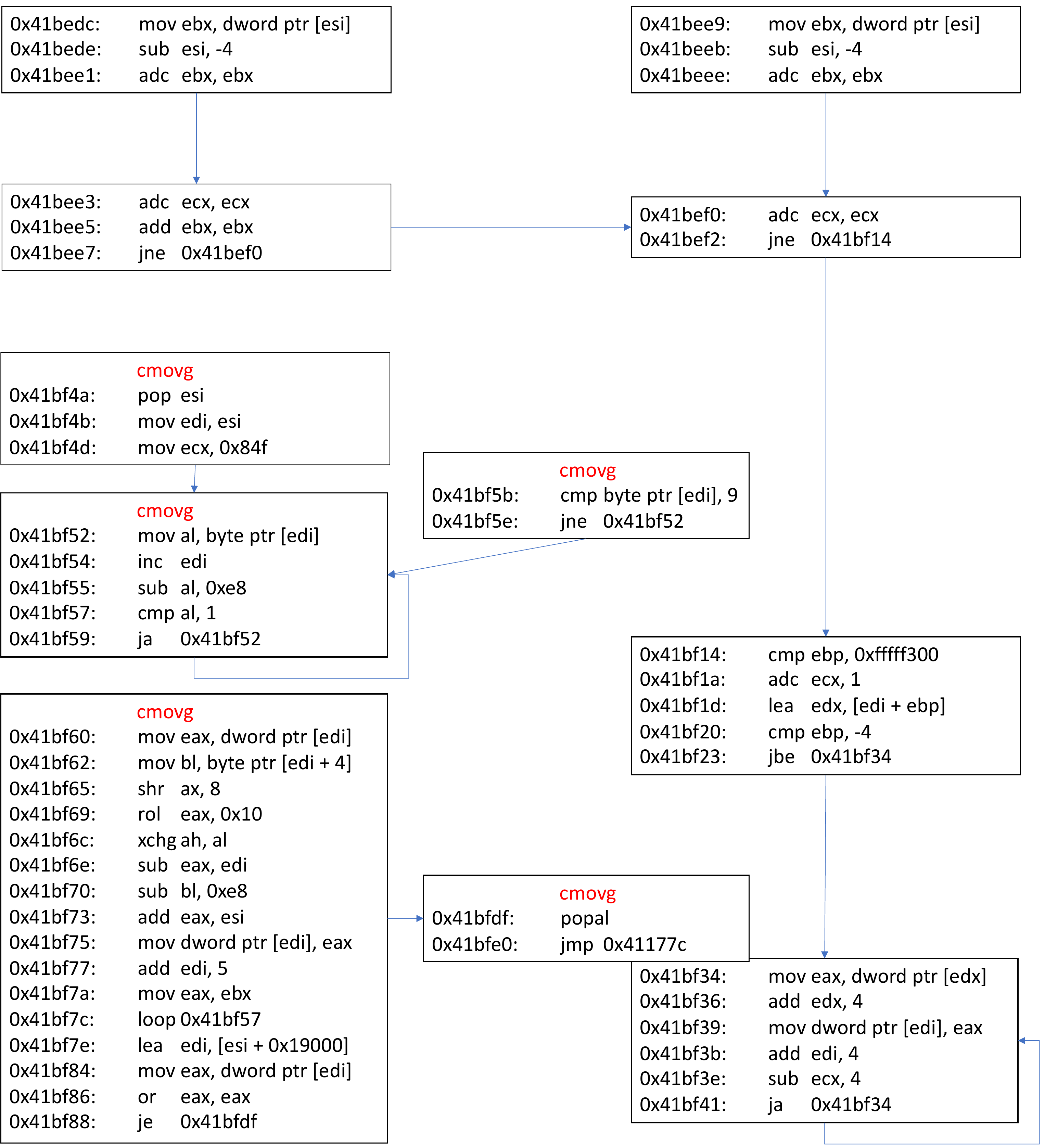}
         \caption{For the DGCNN model}
         \label{fig:sample_gnn}
     \end{subfigure}
        \caption{Injected Features by the SRL attack}
\label{fig:features}
\end{figure}

Figure \ref{fig:features} illustrates the injected features of the proposed attack.
To illustrate the adversarial samples generated by the SRL attack, we take one backdoor CFG (Backdoor.Win32.Assasin) as an example. 
Figure \ref{fig:sample_gcn} and Figure \ref{fig:sample_gnn} are snippets of the sample and the injected features for two models.
These results show that the SRL attack can adapt to the detection models.
For different models, the agent takes different actions on different basic blocks. The agent only requires one semantic {\tt Nop} and its corresponding basic blocks to evade the detection model.

\section{Impact of the injected opcodes}
\begin{figure}[h]
     \centering
     \begin{subfigure}[b]{\linewidth}
         \centering
         \includegraphics[width=\linewidth]{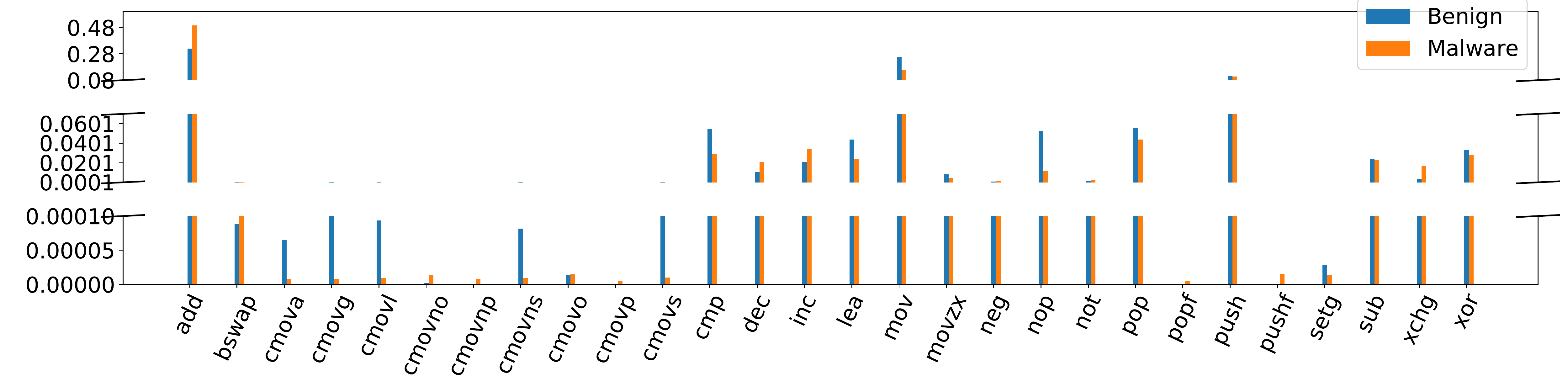}
         \caption{Occurrence frequency of opcodes in the dataset}
         \label{fig:freq1}
     \end{subfigure}
     \vfill
     \begin{subfigure}[b]{\linewidth}
         \centering
         \includegraphics[width=\linewidth]{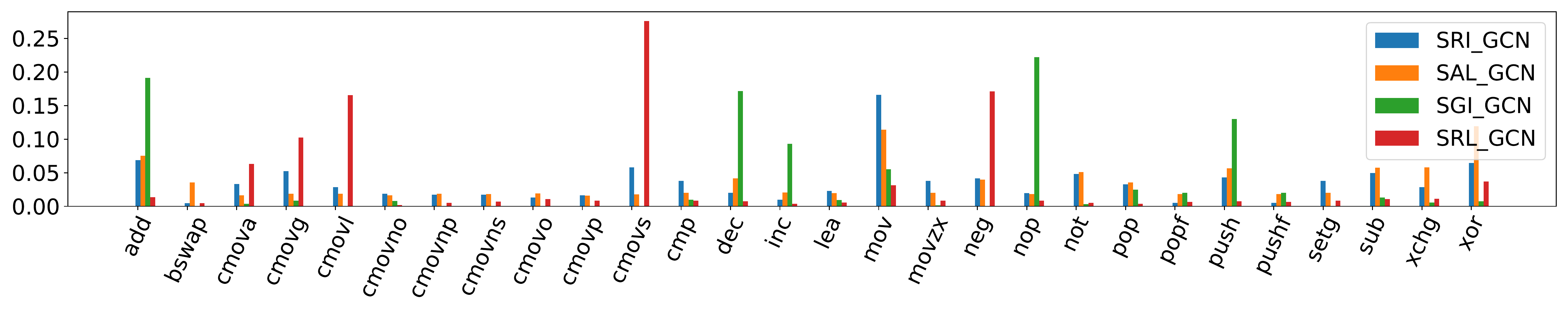}
         \caption{Opcodes chosen by different attacks for the DGCNN model }
         \label{fig:freq2}
     \end{subfigure}
     \vfill
     \begin{subfigure}[b]{\linewidth}
         \centering
         \includegraphics[width=\linewidth]{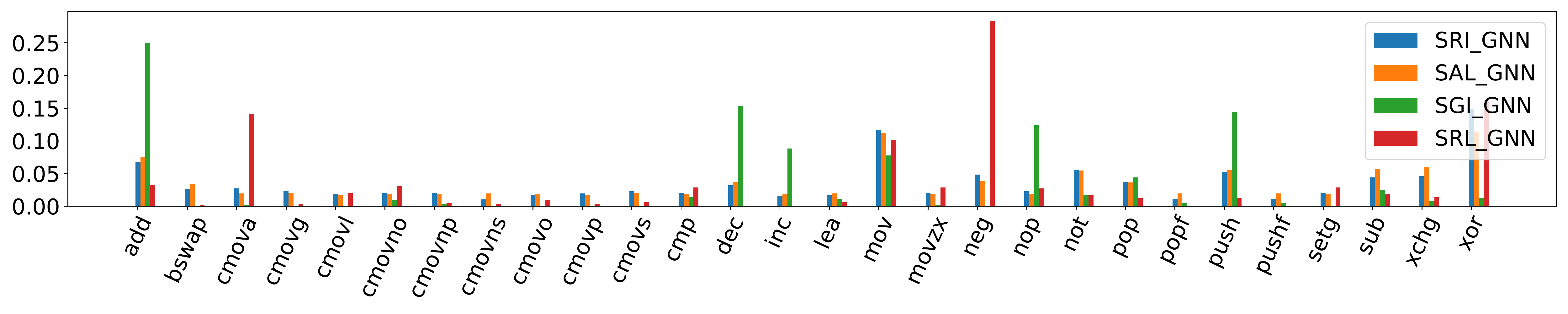}
         \caption{Opcodes chosen by different attacks for the Basic GCN model }
         \label{fig:freq3}
     \end{subfigure}
        \caption{Impact of the inserted opcodes}
        \label{fig:opcode}
\end{figure}

Figure \ref{fig:opcode}  shows the impact of different opcodes.

\end{document}